\documentstyle[aas2pp4]{article}
\begin{document}
\title{Evolutionary stellar population synthesis at 2\AA~spectral resolution}
\author{A. Vazdekis}
\affil{Institute of Astronomy, School of Science, University of Tokyo, 
Osawa 2-21-1, Mitaka, Tokyo 181-8588, Japan \\
e-mail: vazdekis@mtk.ioa.s.u-tokyo.ac.jp}
\journalid{Vol}{Journ. Date}
\articleid{start page}{end page}
\paperid{manuscript id}
\cpright{type}{year}
\ccc{code}
\lefthead{A. Vazdekis}
\righthead{Evolutionary stellar population synthesis at 2\AA~spectral resolution}

\begin{abstract}
In this paper we develop an evolutionary stellar population synthesis model to 
predict spectral energy distributions, SED's, for single-age, single-metallicity 
stellar populations, SSP's, at resolution $\sim$1.8\AA~in two narrow, but very
important spectral regions around 4000\AA~and 5000\AA. The input stellar database 
is composed of a subsample of $\sim$550 stars, selected from the original KPNO 
coude feed stellar spectral library of Jones. Therefore, this is the first time that 
an evolutionary model employs such an extensive empirical stellar spectral library, 
at such high resolution, for supporting its SED's predictions. 

A spectral library corresponding to simple old stellar populations with 
metallicities in the range $-0.7\leq[Fe/H]\leq+0.2$ is presented here, as well 
as an extensive discussion about the most popular system of absorption indices 
at intermediate resolution, the Lick system, showing the advantages of using 
the new model predictions. Also, for the first time is shown the behavior of 
the system of indices of Rose, at higher resolution, as a function of the age 
and metallicity of the stellar population. 

The newly synthesized model spectra can be used to analyze the observed galaxy 
spectrum in a very easy and flexible way, allowing us to adapt the theoretical 
predictions to the characteristics of the data instead of proceeding in the 
opposite direction as, for example, we should do when transforming the observational 
data for using model predictions based on a particular instrument-dependent system 
of indices at a specific resolution. The synthetic SSP spectra, with flux-calibrated 
response curve, can be smoothed to the same resolution of the observations, or 
to the measured galaxy internal velocity dispersion, allowing us to analyze the 
observed spectrum in its own system. Therefore we are able to utilize all the 
information contained in the data, at their spectral resolution. After performing 
this step, the entire observational spectrum can be compared at once, or the 
analysis can be done measuring a particular set of features in the two, the 
synthesized and the observational spectrum, rather than trying to correct the 
latter from broadening or instrumental effects. 
 
The SSP model spectra were calibrated at relatively high resolution with two
well studied metal-rich globular clusters in our galaxy, 47~Tuc and NGC~6624,
providing very good fits and being able to detect well known spectral 
peculiarities such as the CN anomaly of 47~Tuc. The model was also confronted 
to an early-type galaxy, NGC~3379, revealing its well known magnesium over 
iron overabundance, and showing how appropriate are the new model predictions, 
as well as the way in which they can be used, for studying the elemental ratios 
of these stellar systems. 
In fact, different models of different metallicities provide equal approaches 
to the galaxy spectrum: once H$_{\beta}$ is properly constrained, either we are 
able to fit the iron features (with a metallicity somewhat in the range 
-0.4$\leq$[Fe/H]$\leq$0) or the magnesium features (with a metallicity in the 
range 0$\leq$[Fe/H]$\leq$+0.2), but never the two set of indices simultaneously.

\end{abstract}


\section{INTRODUCTION}
The stellar population studies of galaxies, in particular the most simple ones, 
i.e., the early-type galaxies, have been performed originally by using photometric 
information (e.g., broad band filters in the Johnson system) with different types 
of stellar population synthesis methods (e.g., Tinsley 1980). These models try to 
find a combination of stars for which the integrated colors agree with the observed 
ones. This has been done either using a few physical constraints, the so called 
{\it empirical} population synthesis models (e.g., O'Connell 1986), as opposed to 
the {\it evolutionary} which, by means of a theoretical isochrone or H-R diagram 
convert isochrone parameters to observed ones, assuming empirical or theoretical 
prescriptions, and finally integrate along the isochrone assuming an initial mass 
function, IMF, e.g, Arimoto \& Yoshii (1986). These studies showed that there is 
an age-metallicity degeneracy, i.e, the two effects cannot be separated 
simultaneously (Worthey 1994; Arimoto 1996). 

Analyzing the galaxy spectra is possible to obtain more information. This analysis 
can be done by evolutionary stellar population synthesis models which provide spectral 
energy distributions, SED's, at low dispersion, either using theoretical stellar 
atmospheres, e.g., Kurucz 1992 (e.g., Bressan et al. 1994; Kodama \& Arimoto 1997) 
or empirical stellar spectral libraries (e.g., Bruzual \& Charlot 1993). 
Also, on the basis of empirical stellar libraries, SED's predictions are predicted
by empirical population synthesis approaches (e.g., O'Connell 1980; Pickles 1985; 
Bica 1988). In particular, Bica (1988), used as units of population distributions
of stars observed in clusters of our Galaxy, instead of individual stars.

Recently, more accurate spectral information have been addressed by predicting 
absorption line-strengths at intermediate resolution ($\sim$9\AA~FWHM) for the 
strongest atomic and molecular absorption features in the visible, mainly those of 
the extended Lick system (Worthey et al. 1994). These new generation of models 
(e.g., Worthey 1994; Vazdekis et al. 1996, hereafter Paper I) calculate the 
line-strengths of the integrated populations by means of empirical fitting polynomia 
which relate the stellar atmospheric parameters (T$_{eff}$, $\log g$ and [Fe/H]) 
with the measured index equivalent widths (Gorgas et al. 1993; Worthey et al. 1994). 
Some of these absorption indices are also predicted on the basis of other approaches 
than the Lick fitting functions (e.g., Peletier 1989; Buzzoni 1995).
Models of this kind have shown that solar neighborhood elemental ratios are not 
universal in external galaxies (Peletier 1989; Worthey et al. 1992; Vazdekis et al. 
1997, hereafter Paper II). In this paper we combined the information provided 
by almost the whole set of indices of the extended Lick system, together with optical 
and near-infrared broad band photometry. We found the metallicity as the main 
parameter causing the observed color and absorption feature gradients when going 
outward in giant ellipticals. These intermediate spectral resolution indicators 
were found very useful for studying the S0 galaxies (e.g., Bender \& Packet 1995; 
Fisher et al. 1996; Jablonka et al. 1996; Vazdekis \& Peletier 1997; 
Kuntschner \& Davies 1998). This kind of model predictions, in particular those 
based on the Lick system, became very popular and widely used in most of the recent 
works that attempted to understand the stellar populations of the galaxies and 
globular clusters (e.g., Trager et al. 1998 and references therein). However, 
for a correct comparison of these absorption line-strength predictions with the ones
measured on the galaxy data, a number of serious problems must be solved in advance. 
In particular, since the stars of the Lick database are not flux-calibrated, the 
use of the Lick system model based predictions requires a proper conversion of
the observational data to the characteristics of the instrumental response curve 
of this system (see the extensive analysis of Worthey \& Ottaviani 1997, hereafter 
WO97). This is usually done by observing a number of Lick stars, with the same 
instrumental configuration as the one used for the galaxy. Then, comparing 
with tabulated Lick measurements for the same stars, the authors find some empirical 
corrections factors for each individual absorption feature. Another important step 
to be followed, is to prebroaden the observational spectra to match the resolution 
of that system (i.e., in most cases neglecting the higher resolution allowed by the 
original data), which is strongly dependent on the wavelength. From this step, 
a new set of empirical corrections factors for each individual feature is 
found on the basis of the observed common stars. Finally, the authors place 
their line-strength measurements on to the Lick system by applying these two 
types of correction factors. Thus, the analysis of the observed galaxy spectrum requires 
a very careful work, and do not allow to achieve all the potential offered by the 
data because the fixed requirements of the adopted system. 

The higher the spectral resolution the higher the constraining power. However 
predicting such high dispersion SED's for stellar populations is very difficult 
due to the non availability of the required input stellar spectra. Among the 
problems the theoretical stellar spectra require heavy and non-available
calculations while the empirical ones usually do not cover all the desired 
atmospheric parameters. There are a few studies that have attempted to include 
spectral features at higher resolution (e.g., Rose 1994, hereafter R94; Jones 
\& Worthey 1995). The first author used empirical stellar spectra and 
Tripicco \& Bell (1992, 1995) used theoretical model atmospheres to synthesize 
stellar population spectra for analyzing metal-rich globular clusters providing 
acceptable fits. R94 showed that the answer obtained from high spectral resolution 
analysis would be different from the one obtained by, e.g., taking into account
colors. 
Despite the fact that 47~Tuc and M~32 show very similar colors, he 
found significant differences in their stellar populations on the basis of higher
resolution spectra. Jones \& Worthey (1995) showed the potential of their 
H$\gamma_{HR}$ high resolution index to separate the metallicity and the age 
effects. WO97 have shown the importance of the increasing resolution when working 
with Balmer lines.

In Paper II we showed the necessity of using as much observational constrains as 
possible for increasing our discriminating power between the number of possible 
solutions obtained when trying to fit early-type galaxies by only taking into account
colors. With the new model developed here we hope to increase even more our 
constraining power synthesizing spectral energy distributions in 
two narrow wavelength ranges around 4000\AA~({\it blue}) and 5000\AA~({\it red}) 
but at higher resolution ($\sim$1.8\AA~FWHM). This method is different from the
previous studies because is the first time where SED's at such resolution are 
predicted under the evolutionary stellar population synthesis machinery 
on the basis of an extensive empirical stellar spectral library of Jones (1997, 
hereafter J97). This model differs, e.g., from the one of Jones \& Worthey (1995) 
in, e.g., the fact that SED's corresponding to single-age single-metallicity
stellar populations (SSP's) are calculated, instead of predicting just a number 
of features at relatively high resolution via the use of empirical fitting 
functions of the type of those calculated by Worthey et al. (1994). It also 
differs from the model of Tripicco \& Bell (1995) because uses an extensive 
empirical stellar spectral library, instead of theoretical stellar atmospheric 
calculations. 

Section~2 explains the main points regarding the empirical stellar spectral 
library to be suitable for the stellar populations synthesis purpose. 
Section~3 explains in detail how our population synthesis model (Paper I) has 
been developed for synthesizing SED's at 1.8\AA~FWHM for old SSP's. 
In section~4 the intermediate resolution indices of the Lick system are measured 
on the synthetic spectra and compared with previous model predictions, showing the 
advantages of the use of the new SSP spectral library. For the first time is
shown the behavior of the high resolution indices introduced by R94 as a 
function of the metallicity and age of the stellar population. 
In section~5 the model spectral library is calibrated with observational 
data of two well known metal-rich globular clusters. We discuss how these 
predictions can be used for analyzing the data. 
In section~6 we test the potential utility of the new model spectra and how they 
must be applied when studying an early-type galaxy.
Finally, in section~7, the conclusions are presented.

\section{Treatment of the empirical stellar spectral library}
\subsection{The stellar library}
In this paper our evolutionary population synthesis model (Paper I) is extended on 
the basis of the empirical stellar library of J97. The stars were selected to cover 
the most important evolutionary phases of a wide variety of spectral types (O - M), 
luminosity classes (I - V) and metallicities (-2.5$\leq$[Fe/H]$\leq$+0.5), on 
the basis of the availability of their atmospheric stellar parameters. 
This library is composed of 684 stars for which the spectra were obtained using 
the coude feed telescope and spectrograph at KPNO with grating RC-250 and the 
TI5 800$\times$800 pixel CCD. The spectra cover two windows, 3820-4500\AA~and 
4780-5460\AA, at a resolution of 1.8\AA~FWHM. The reader is referred to J97 and 
Leitherer et al. (1996) for more details about the library.
In the present paper these two spectral ranges will be called {\em blue} and 
{\em red}, respectively. These two wavelength regions provide a large number 
of well studied absorption line indices: 14 of the 21 absorption features of 
the extended Lick system (Worthey et al. 1994), the new Balmer indices of
WO97, the high resolution indices introduced by R94, and 3 Lick-style Balmer 
indices by Jones \& Worthey (1995).

Since the purpose here is to use this empirical library as a stellar spectral 
database for our stellar population synthesis model, those stars which were  
peculiar were removed from the original sample. In fact, 137 stars were removed 
from the sample because the following reasons: spectroscopic binaries, those stars 
that, after checking their spectra, were found to be quite different from what 
expected on the basis of the given atmospheric stellar parameters and those stars 
with high signal of variability (except for very late or very early spectral types). 
However, for some parametrical regions, for which only a few stars are present,
we have chosen to keep some of these anomalous stars for completeness, e.g., very 
hot or very low metallicity stars with [Fe/H]$<$-1.0.

\subsection{Reddening}
To account for the reddening affecting the colors of these stars we found in the 
literature around $\sim350$ for which extinction values were given by different 
authors. In particular we are very greatful to V. Vansevi\^cius (private 
communication) who provided the E$_{B-V}$ values for the largest number of stars. 
Other papers from where we took large number of extinction values were 
McClure (1970), Neckel et al. (1980), Bond (1980), Savage et al. (1985), 
Beers et al. (1990), Blackwell et al. (1990), Carney et al. (1994), 
Blackwell \& Lynas-Gray (1994) and Alonso et al. (1996). Finally, we assumed
zero extinction for nearby stars, closer than $\sim50pc$, for which we did not
find any value in the literature.

\subsection{Absolute magnitude}
The absolute magnitude for each star was derived using the paralaxes measured
by Hipparcos. Because the stellar fluxes are going to be assigned by our
population synthesis model (this point is extensively discussed in section~3),
the derived distances were kept even for the farthest stars to have
an estimation for the M$_{V}$ as a reference, rather than obtaining an
accurate value. For a small number of stars for which no paralaxes were given, 
we either used the absolute fluxes of Bond (1980), V. Vansevi\^cius (private 
communication), or we just assigned the expected values following their given 
atmospheric parameters.

\subsection{Adopted stellar atmospheric parameters}
For each star of the sample we searched in the literature for a number of sets 
of atmospheric stellar parameters (T$_{eff}$, $\log g$, [Fe/H]) provided by 
different authors. The main sources were J97, Cayrel et al. (1997), 
Edvardsson et al. (1993), Carney et al. (1994), Alonso et al. (1996), 
Worthey et al. (1994), Borges et al. (1995), Pilachowski et al. (1996), 
Axer et al. (1995) and Bonifacio \& Molaro (1997). On the basis of these 
different sets of stellar atmospheric parameters we derived the corresponding 
B-V and V-I colors following the prescriptions given in Paper I (see also 
Section 3), and then compared with the observed ones after dereddening (using
the values found in section 2.2). The set of atmospheric parameters that provided 
the smallest sum of squared residuals for these two colors were selected. However 
for those stars for which no extinction values were given we just kept the most 
recent values given by Cayrel et al. (1997) or by J97. At this step, we removed
those stars for which the predicted colors were very different from the 
observed ones. In conclusion, for each of the remaining stars we kept the 
adopted T$_{eff}$, $\log g$, [Fe/H], M$_{V}$ and the corresponding blue and 
red spectrum, as a database for our stellar population synthesis code. 

The adopted parameters for this final subsample composed of 547 stars are plotted 
in Fig.~1. The full list of stars and parameters can be found in Vazdekis (1998). From 
these diagrams it can be seen that all type of stars are well represented for solar 
metallicity. Normal giants with metallicities in the range $-0.7\leq[Fe/H]\leq+0.4$ 
are also well covered. However this is not the case for giants cooler than 
$\sim$3800~K, where mainly solar metallicity stars are present. 
The situation is the same for dwarfs cooler than $\sim$5000~K and hotter than 
$\sim$6800~K. Unfortunately, we do not see many metal-rich dwarfs with
metallicities larger than [Fe/H]=+0.2, preventing to build very high metal-rich 
models. Finally, we see that there is an acceptable surface gravity coverage. 
In general, we can conclude that the 
metallicities for which we are safe are in the range $-0.7\leq[Fe/H]\leq+0.2$. With 
respect to the limitations arising from the temperature coverage of the stellar 
sample, we will limit the present models to cover ages larger than 6~Gyr for 
[Fe/H]=-0.7, 3~Gyr for [Fe/H]=-0.4 and 1.5~Gyr for models of higher metallicities. 

\subsection{The flux-calibration quality}
In order to check the flux-calibration quality of the stellar spectra of J97, 
we asked Cardiel \& Gorgas (private communication) to provide us with some of 
their intermediate resolution, but very well flux-calibrated, stellar spectra 
of the stars in common with the J97 library. Twelve of these $\sim$120 common 
stars were selected, taking into account the fact of having different spectral 
types and air masses at the time of observation. 
Next, four well separated spectral regions, covering $\sim$30\AA~each, not 
very crowded by absorption features were defined for the red and blue spectra. 
For this purpose the pseudocontinua spectral regions given by Worthey et al. 
(1994) were taken into account. Then the stellar spectra of J97 were smoothed 
to match the Cardiel \& Gorgas resolution. Next, the fluxes at these spectral 
intervals were measured in order to derive colors for the two sample of stellar 
spectra. The comparison of these colors showed that the J97 flux-calibration 
quality is reasonably good when the colors are composed from nearby spectral 
regions. 
However for larger wavelength baselines colors, i.e. composed of spectral
regions separated by more than $\sim$300\AA, we obtain differences that can
be slightly larger than 0.1 magnitudes. Therefore this test confirms that the
the stellar spectra of J97 are nearly flux-calibrated, and we conclude that the 
flux-calibration quality is acceptable within $\sim$300\AA. 

\section{The model}
This section explains how our evolutionary stellar population synthesis model (Paper I)
is used for calculating SED's, at relatively high spectral resolution, for SSP's on the 
basis of the empirical stellar library described in section~2. The reader is referred to 
Paper I for a complete model description, however we briefly summarize here its main aspects. 
The model makes predictions for the optical and IR colors and 29 absorption line indices of 
the extended Lick system (Worthey et al. 1994; WO97) and the Ca{\sc ii} triplet features in 
the near infrared as defined by D{\'{\i}}az et al. (1989). It uses the homogeneous set of 
the theoretical isochrones of Bertelli et al. (1994) (calculated with solar abundance ratios) 
and the stellar tracks of Pols et al. (1995) for the very low-mass stars. We converted to 
the observational plane (e.g., fluxes; colors) on the basis of the relations inferred
from extensive observational stellar libraries rather than by implementing theoretical
stellar atmospheric spectra. In particular, for normal dwarfs we used the metal dependent 
empirical relations given in Alonso et al. (1996) (obtained from their sample composed of 
$\sim$500 stars) to obtain each color as a function of the $T_{eff}$ and metallicity. 
For stars above 8000 K we performed the $T_{eff}$-color transformation of Code et al. 
(1976) and the color-color relation from Johnson (1966). For normal giants we used the 
empirical calibration of Ridgway et al. (1980) to obtain the V-K color from 
$T_{eff}$, after which we applied the color-color conversions of Johnson (1966) (for giants) 
and the empirical stellar library of Fluks et al. (1994).
For the very cool giants we used the models of Bessell et al. (1989, 1991) and the Fluks 
et al. (1994) empirical stellar library. Finally, for predicting the line-strengths of the 
Lick system of indices, the model uses the empirical fitting functions obtained by Worthey 
et al. (1994) and Gorgas et al. (1993), while for predicting the Ca{\sc ii} triplet 
features we produced our own empirical fitting functions. As can be inferred from the above 
description, this model is an evolutionary stellar population synthesis code which employs, 
as far as possible, extensive observational stellar libraries rather than theoretical 
stellar atmospheric spectra to support its photometric predictions. This fact 
differenciates this model from most of the present day evolutionary stellar 
population synthesis models, which are heavily based on, e.g. the theoretical stellar 
spectra of Kurucz (1992), from which the colors are calculated. The code is also able 
to produce more complex stellar populations if we use its full chemo-evolutionary option 
(see Paper I). 

Despite the fact that this model provides spectral information in the form of 
line-strengths, there is a lack of SED predictions. In this paper the model is
extended to predict SED's at two narrow spectral regions but at 1.8\AA~FWHM resolution, 
always keeping our criterion to give the priority to the use of the empirical stellar 
libraries.
Following, we explain how to make such predictions starting at the point when the
code yields a stellar distribution for an SSP of age $T_{G}$ and metallicity $Z$ 
(we refer the interested readers to Paper I for details regarding previous steps 
of the model calculation). To obtain the high resolution spectrum corresponding to 
a SSP, $S_{\lambda}(T_{G},Z)$, we integrate in the following way:

\begin{equation}
S_{\lambda}(T_{G},Z)=\int_{m_{l}}^{m_{T_{G}}}S_{\lambda}(m,T_{G},Z)F_{\lambda_{ref}}(m,T_{G},Z)N(m,T_{G})dm
\end{equation}

\noindent
where $S_{\lambda}(m,T_{G},Z)$ is the empirical spectrum corresponding to a star of 
mass $m$ and metallicity $Z$ which is alive at the age assumed for the stellar 
population $T_{G}$, $F_{\lambda_{ref}}(m,T_{G},Z)$ is its corresponding absolute 
flux at certain wavelength reference point, and $N(m,T_{G})$ is the number of this 
type of stars. The spectrum to be assigned to each of these stars is selected from 
the empirical stellar database, while the corresponding total flux is assigned 
following the prescriptions of our code. Therefore we need to find a common reference 
wavelength point, $\lambda_{ref}$, from where to scale S$_{\lambda}(m,T_{G},Z)$. 
We selected a $\sim$30\AA~wide spectral range, not very crowded by absorption lines 
along the stellar spectral type, around the peaks of the B and V bands filters. 
These regions are (4421-4451\AA) for the blue, and (5415-5443\AA) for the red. 
The two regions fall almost at the red edge side of each spectrum (blue and red). 
We divided the stellar spectra of the whole empirical stellar sample by the average 
value at the selected regions, so that we obtain a value of 1 at 4436\AA~ for the blue 
and at 5429\AA~ for the red. To calculate the absolute flux at the selected reference 
points, $F_{\lambda_{ref}}(m,T_{G},Z)$, for each of the stars requested by the code, 
we first used equation (6) of Code et al. (1976), which provides a relation between 
the absolute V flux and the calibrated bolometric correction, to infer the following 
relation:

\begin{equation}
\frac{F_{V}}{F_{bol_{\odot}}} = 10^{-0.4(M_{V}-3.762)}
\end{equation}
\noindent
where $F_{V}$ is the absolute V flux, and the absolute magnitude is provided by the  
code. The predicted model color B-V is used to calculate the absolute B flux by:

\begin{equation}
\frac{F_{B}}{F_{bol_{\odot}}} = \frac{F_{V}}{F_{bol_{\odot}}} 10^{-0.4(B-V)}
\end{equation}

\noindent  
Finally, the absolute V and B fluxes are divided by their effective equivalent widths
to obtain the absolute flux per Angstrom, in order to be assigned to the two selected
reference wavelength points. 

The empirical stellar spectrum $S_{\lambda}(m,T_{G},Z)$, to be assigned to a 
requested star for which the model provides T$_{eff}$, $\log g$, [Fe/H] and M$_{V}$, 
is selected from the whole sample of 547 stars according to the following method:

\begin{itemize} 
\item Depending on whether the required star is giant ($\log g < 3.5$) or dwarf 
($\log g\geq 3.5$), the model selects from the sample all these stars with temperatures 
in the range (T$_{eff}-\Delta$T$_{eff}$, T$_{eff}+\Delta$T$_{eff}$). 

\item Next, the remaining stars are discriminated taken into account their metallicities,
which must be in the range ([Fe/H]-$\Delta$[Fe/H], [Fe/H]+$\Delta$[Fe/H]). 

\item In the last step, the code selects those stars that simultaneously satisfy 
the conditions of having $\log g$ and M$_{V}$ in the ranges 
($\log g$-$\Delta \log g$, $\log g$+$\Delta \log g$) and
(M$_{V}$-$\Delta$M$_{V}$, M$_{V}$+$\Delta$M$_{V}$), respectively.

\end{itemize}
As can be seen above, the first criterion taken into account was the effective 
temperature and the condition of giant or dwarf. The metallicity was the second 
constraint applied. The final step is performing a more fine tuning for the
luminosity class. The subsequent step is only followed  when having found at 
least one star from the sample after applying the present step. At a certain step,
when no stars are found, the range is gradually increased in a much more fine 
tuning until at least one star is found. In spite of the fact that our starting
minimum box is very small (of the order of typical error stellar atmospheric 
parameter determinations) in most of the cases, after this selection procedure, 
remain more than just a single star because our stellar database is quite large. 
Therefore, usually, the code averages the surviving stellar spectra. However, 
in no case is performed any interpolation between stars with different 
atmospheric stellar parameters. Bruzual \& Charlot (1993) showed that only in
5$\%$ of the cases the situation improved when interpolating (also, their stellar 
library is in fact around 5 times smaller than ours). When dealing with a 
parametrical region not covered by any star, the code chooses the closest one. 
For the most important cases the minimum box described above was: 
$\Delta$T$_{eff}$=90 K, $\Delta$[Fe/H]=0.13, $\Delta \log g$=0.6 and 
$\Delta$M$_{V}$=0.75. Of course, this selection worked very well for the solar 
metallicity as inferred from Fig.~1. However, to account for the fact that always 
are found more number of stars at the half part of the box which is closer to 
[Fe/H]=0.0, for [Fe/H]=-0.7 and [Fe/H]=+0.2 (see below) we left our starting 
metallicity ranges asymmetrical, being the less crowded side two times greater. 
Also more generous ranges were established for those parametrical regions where 
not many stars can be found (e.g., for cool and hot stars for metallicities 
different than solar). In this way we avoided the fact to, e.g., assign a very 
hot metal-poor star to a metal-rich one, because the T$_{eff}$ was the first 
criterion applied. This does not mean that the order of the steps explained 
above is changed, since the accuracy of the given temperatures, e.g., above 
7000~K, is in fact much lower.

\paragraph{Models of metallicity [Fe/H]=+0.2.} 

The present stellar spectral library does not allow us to build very high 
metallicity models. However, since our effort is mainly devoted to study 
galaxies it is worth to show models for as higher metallicity as it allowed. 
Due to the fact that the limit imposed by our stellar 
library is [Fe/H]=+0.2 (see section 2.4) and because our stellar population 
synthesis model is based on the Padova isochrones (calculated for solar and 2.5 
times solar metallicity, i.e., [Fe/H]=0.0 and [Fe/H]=+0.4, respectively), we 
need either to interpolate between the isochrones before using them for 
calculating the SSP synthetic spectrum, or to interpolate the final output 
spectra after calculating them on the basis of the closest isochrones. 
For simplicity, we followed the second approach: for each assumed age, we first 
calculated the spectrum for an SSP corresponding to [Fe/H]=+0.2 on the basis 
of the isochrones of metallicity [Fe/H]=0.0, and another spectrum on the basis 
of the isochrones of [Fe/H]=+0.4. Finally, the two resulting spectra are combined 
to obtain one which resembles a metallicity of [Fe/H]=+0.2. In fact we find
this way safe enough since our method of selecting the stellar spectra is based
on the desired metallicity ([Fe/H]=+0.2) rather than on the metallicity of the
isochrones. Also, in order to ensure the participation of a larger number of
metal-rich stellar spectra we selected an asymmetrical starting box in the
metallicity range $+0.1\leq [Fe/H] \leq +0.4$. Thus, all the stars involved
in the two calculations do have metallicities around [Fe/H]=+0.2, while the 
main differences between the two resulting spectra arise from the fact 
that the average effective temperature of the SSP spectrum obtained on the 
basis of the solar metallicity isochrone is somewhat $\sim$200~K higher 
(if we average between all the ages above 1.5~Gyr).

\section{The synthetic SSP spectra at 1.8~\AA}
Instead of giving an extended list of the new model spectra and/or tables of 
different spectral indices as measured on the resulting spectra, in this paper 
we have preferred to study the behavior of the Lick system of indices, at 
intermediate resolution, when compared with model predictions based on the 
empirical fitting functions derived from the stellar Lick library (see section~1). 
Here we also show for the first time the behavior of the higher resolution indices 
of R94 as a function of the metallicity and age. For this purpose, synthetic spectra 
at 1.8\AA~FWHM (0.6\AA/pix) corresponding to single-aged simple stellar populations 
of metallicities [Fe/H]=-0.7, -0.4, 0.0 and +0.2 were calculated. For each metallicity, 
the ages were varied from 1.6 to 17~Gyr, except for the lowest metallicity 
where we started at 6.3~Gyr and for [Fe/H]=-0.4 starting at 3.2~Gyr. The reason for
not calculating younger ages arises from the limitations of the present library
as already discussed in section~2.4. To be more general, in this section we used 
the Salpeter (1955) initial mass function. The resulting spectra cover the 
following wavelength ranges: 3856-4476\AA~for the blue and 4795-5465\AA~for 
the red.

The extended data set including the variation of the IMF, as well as a discussion 
regarding the behavior of these spectral indices at different resolution, i.e., at 
different galaxy velocity dispersions, will be published elsewhere.  

\subsection{The Lick indices}
This section shows a comparison between the Lick indices as measured on the 
synthetic spectra and as calculated in Paper I, i.e., on the basis of the 
empirical fitting functions of Worthey et al. (1994) and Gorgas et al. (1993). 
However, transforming our line-strength measurements on the new spectra to 
the Lick system (i.e., the predictions in Paper I) is not so straightforward, 
because this system is based on an stellar library which is not flux-calibrated, 
and because its resolution FWHM depends on the wavelength range (see WO97 for 
an excellent review). Also, these authors tested very carefully, various methods 
with different stellar libraries, yielding to a large scatter regarding the 
resolution. Here, for simplicity, we prebroadened our SSP's synthetic spectra by 
convolving with gaussians (see Table~2) which allowed us to match our measurements 
on the smoothed spectra for two strongly resolution sensitive indicators, the 
Ca4227 in the blue and the Fe5335 in the red, to the model predicted values 
calculated on the basis of the empirical fitting functions for the solar 
metallicity models. Table~1 shows the measured values for different 
metallicities and ages, while Table~2 provides the way for converting the new 
index values to the previous predictions of Paper I, i.e., on the Lick system
instrumental response curve and resolution.  

We perform here a comparison on absolutely similar model prescriptions (the 
code is the same one), representing, in fact, a comparison of the Lick indices 
for SSP's as derived from the Lick instrumental response curve and as measured 
on the SSP's flux-calibrated spectra. The main difference is the different
input stellar libraries, however, since the number of stars is quite large 
in the two samples the discussion presented here cannot be significantly 
affected. The results are summarized in Fig.~2. 

As it was noted above, the Ca4227 and Fe5335 line-strengths are fully matched 
for solar metallicity. Good agreement is also found for the resolution sensitive 
indicators Fe5270 and Fe5406. 

The Mg$_{1}$ and the Mg$_{2}$ show important differences which become larger as 
a function of the increasing metallicity.  We attribute this to the fact that in 
their index definitions the two pseudocontinua are separated by around 470\AA, 
being in this way very sensitive to the differences between the spectrum shape 
of our nearly flux-calibrated (very safe within $\sim$300\AA~but smaller than the 
required range) model spectra and the instrument dependent Lick indicators, and 
because the characteristic Mg bump becomes more prominent as a function of the 
metallicity. The same effect is seen for the much narrower Mg$_{b}$, but to 
a much less extent. 

The observed model shifts for the CN are surprisingly small despite the fact 
that WO97 showed a rapid decrease of the resolution FWHM of the Lick system as 
a function of decreasing wavelength for the regions corresponding to our blue 
spectra, and because the large wavelength coverage of the CN indices. This fact 
was already pointed out by these authors on the basis of their comparison of
various stellar samples. 

The long wavelength coverage definitions must contribute in part to the observed 
differences for the Fe4383 and Fe5015. The latter also shows an increasing 
shift for a decreasing metallicity. However, in this case, the trend depends to 
some extent to the resolution adopted.
The model differences for Fe4383 and Fe5015 are somewhat lower and higher, 
respectively, than the shifts given by WO97 in their Table~9. These authors
performed a direct comparison of the spectra of the stars in common between the 
Lick and Jones samples, after broadening the latter stellar sample by their favorite
values provided in their Table~8. These differences must be attributed in part to
the differences in smoothing, i.e., if the applied broadening is quite different,
this trend could well be reversed.   

The behavior of the G band is slightly different from what expected from 
the calculations based on the empirical fitting functions. The G band tends to 
converge to the same level for old populations at different metallicities. Notice 
that, for this index, the dependence of the differences between the two kind of 
predictions increase with decreasing metallicities (in opposite direction to the 
Mg$_{1}$ and the Mg$_{2}$ indices). However, as for the Fe5015, this is in part
determined by the applied smoothing. 

Finally, the new strengths obtained for H$_{\beta}$ 
are slightly higher than those based on the empirical fitting functions for the
oldest populations. The model differences are larger for increasing metallicity 
but the average value is very similar to the one given in WO97.  

In Fig.~3 we perform the same kind of comparison as in Fig.~2 for the new Balmer 
indices defined in WO97. In this plot we have included their model predictions for 
solar metallicities. In general we see that our values, either as calculated on 
the basis of their empirical fitting functions or as measured on the new spectra, 
are lower than their predictions. These differences are attributed mainly to the 
fact that our respective models are based on different set of isochrones, being 
ours slightly cooler. The same explanation was given for the observed differences 
in H$_{\beta}$ for the model comparison performed in Paper I. 

After analyzing Fig.~2 and Fig.~3 we should warn the users of the stellar 
population synthesis models to be aware of these differences, which can lead
to different results when interpreting the observational data. For example, if 
we want to use the Mg$_{2}$ or H$_{\beta}$ vs. age diagrams for interpreting some
observed values that, e.g., fall onto a $\sim$12~Gyr solar metallicity for the
model predictions based on the empirical fitting functions (see thin solid line on 
Fig.~2), it can also be interpreted as a result of a stellar population of the
same metallicity but $\sim$17~Gyr old when using the flux-calibrated SSP model
spectra (thick solid line). The situation can be even worse if the observed 
values are not properly transformed to the characteristics of the Lick system 
instrumental response curve and resolution(s). 
In this case, model predictions based on the empirical fitting functions can 
drive easily to erroneous interpretations. Also, Fig.~2 reveals to what extent 
can be difficult to perform a proper data conversion to the Lick system since
the response could also depend on model parameters such as the metallicity or age.
In fact, when looking at, e.g., the Mg$_{2}$ plot, we see that a correct 
conversion to the Lick system cannot be performed without knowing {\it a priori} 
the metallicity of the observed target which, in fact, is intended to be studied. 
It is rather clear that the model spectral library presented here represents a 
considerably more flexible tool, and overcomes most of the previous problems, when 
using the Lick indices for an easier and correct stellar population interpretation.
For increasing the analyzing power, the interested users should first flux-calibrate 
their observational data or normalize in the same way the two, the observed and 
the synthetic SSP model spectra. Next, by smoothing the theoretical SSP spectra 
to the galaxy velocity dispersion or data resolution, we will be able to treat 
each galaxy spectrum in its own system and interpret the information at the highest 
spectral resolution allowed by the data. The present approach prevents for 
performing the heavy task of correcting the particular indices measured on the 
galaxy spectrum for the broadening and instrumental response curve effects, as it 
is required when applying the Lick-system based stellar population model predictions. 
Finally, if our purpose is to compare a group of galaxies of different internal 
velocity dispersions under absolute similar resolution conditions, then we always 
are able to smooth all the data as well as the model spectra to a common velocity 
dispersion value.   
Examples of how to use the new SSP spectral library and this approach for performing
the analysis is given in sections 5 and 6.

\subsection{Rose indices}
Since our SSP's spectra can be obtained for higher resolution than the one(s) 
required by the Lick system, it is worth to show 
the behavior of the R94 system of indices with model parameters. This system
has the advantage to contain many indicators which are in some cases dominated 
by different elements from those of the Lick system, showing important 
dependences on ages, metallicities and luminosity type in a very reduced 
portion of the galaxy spectra around 4000\AA. The way in which these 
indices have been defined differs considerably from the Lick style indicators. 
We refer the reader to R94 (and references therein) where the author has 
extensively discussed and applied this high resolution system of indices 
to observational data of globular clusters and galaxies. 

Fig.~4 shows the dependence of these indices as a function of the metallicity 
and age of the stellar population. The indices were directly measured on the 
resulting spectra, without performing any degradation of the resolution (i.e., 
1.8\AA~ FWHM). Their dependence on the resolution will be fully addressed in 
another paper and in Vazdekis \& Arimoto (1998). The three indices 
Fe{\sc I}$_{HR}$, Ca{\sc I}$_{HR}$ and H$\gamma_{HR}$ were measured following 
the prescriptions given in Jones \& Worthey (1995) who sligthly changed the 
original definitions of R94. 

In general, we see that most of the indices evolve smoothly in spite of the 
fact of the high resolution involved. Most of the indices behave like most of 
the Lick indices: higher values for higher metallicities and ages. 

On the other hand, the 4384/4352 and H$\gamma_{HR}$ indices decrease with 
these two model parameters, however, the latter is much less affected by 
the metallicity effect around solar values. H$\gamma_{HR}$ has been claimed by 
Jones \& Worthey (1995) as the strongest age discriminator. In particular they 
find that around 12~Gyr its derivative $d\log t / d\log Z$ is zero. We find 
that its dependence on the metallicity is not very important but is not 
negligible depending on the metallicity level. In fact their finding is fully 
true for solar and half a solar metallicity models. It is worth to note that 
these authors smoothed their spectra by 83~kms$^{-1}$. However we see a strong 
dropping of the values for models with [Fe/H]=-0.7. Also, our most metal-rich 
models of $\sim$17~Gyr show very similar values to those of solar models of 
$\sim$11~Gyr. A redefinition of this index by using the new SSP spectral library
allowed us to increase substantially its already high discriminating power
(Vazdekis \& Arimoto 1998). In that paper we also address the problem of its
behavior as a function of the resolution. 

Finally, the remaining two indices, Ca{\sc II} and p4220/4208, are nearly constant 
within the covered range of ages despite the fact that the scale is considerably narrow. 
For a detailed discussion about these indices and their errors we refer the reader 
to R94 (and references therein).  

\section{Calibration with globular clusters}
In principle, the globular clusters represent uniform stellar populations and 
therefore they can be used to test the new synthetic model spectra corresponding 
to single-age, single-metallicity populations. Two metal-rich clusters in our
galaxy, 47~Tuc and NGC~6624, were selected for this purpose. J. Rose kindly 
provided his blue flux calibrated spectrum of 47~Tuc and the non flux calibrated 
one for NGC~6624, both at $\sigma \sim 83~Km s^{-1}$. Spectra that cover the two
ranges analyzed in this paper at resolution 3.4\AA~FWHM were generously provided
by S. Covino. We also used a flux-calibrated spectrum at 8.1\AA~FWHM covering the
two spectral ranges for 47~Tuc and a non flux-calibrated one covering the red 
region at 2.1\AA~FWHM for NGC~6624, the two spectra provided by S. Covino.
On the other hand, for performing a comparison between these observations and our 
predictions we have preferred to calculate model spectra using the bimodal IMF as 
defined in Paper I, which only differs from the Salpeter IMF in the fact that 
it decreases the number of stars with mass lower than 0.6M$_{\odot}$, 
while its slope is the same. This selection is favored because the present 
stellar database does not contain many of these stars, and since we showed in 
Paper II that it provided slightly better fits for a small number of standard
early-type galaxies.
Following, we provide two examples of how to study the collected data. First,
we deal with the whole information contained in the spectrum at once. Next,
we study some key individual spectral features at the highest resolution allowed
by the data.

The height of the resulting peak when crosscorrelating the observed and synthesized 
spectra provides a rough but an overall overview of the similarities between them.  
Also, the height of this peak only provides us a relative idea of the goodness of 
the global spectra as they are, but not an absolute value when comparing the obtained
peak heights between different objects or observations. To prepare the model 
spectra for this purpose, they were conveniently degraded by convolving with a 
gaussian function to match the resolution of the observed data. Then, the 
observational and the model spectra were rebinned logarithmically and normalized
using a spline3 of a fixed low order to remove the continua. The continuum removal 
was performed with great care taking into account the peaks of the spectra. When 
the observed spectra would have been flux-calibrated, this task was considerably easier 
and safer. However if the spectrum still contains the instrumental response, large
scale structure can be very difficult to remove and cause not only the lose of the 
information contained in indices with large wavelength coverage, but also narrower 
indicators contained within larger scales can be affected during the normalization 
step. Prior to the crosscorrelation, the spectra were adequately filtered to remove 
the highest frecuencies and multiplied by a cosine-bell-like functions (see Tonry 
\& Davies 1979 and Gonz\'alez 1993 for details about this method). 
As an example of this application, Fig.~5 shows the height of the peak obtained after 
crosscorrelating the blue spectrum of 47~Tuc with model spectra of different 
metallicities and ages at $\sigma \sim 83~Km s^{-1}$. One sees a clear cut difference
between the solar and the models of metallicity [Fe/H]=-0.7. A very similar plot is 
obtained when performing this test using the data provided by Covino at 3.4\AA~FWHM and 
the SSP spectral library prebroadened to match that resolution for the blue and
red spectral ranges. Finally, the same results are obtained for the spectra 
at 8.1\AA~FWHM, in spite of the fact that this metallicity cut is less pronounced.
For saving space we do not show any of these plots, as well as those performed for
NGC~6624, for which the same results arise even when the red spectrum at 2.1\AA~FWHM
is involved.
Fig.~5 suggests that the most likely synthetic spectrum should be found among models 
of [Fe/H]=-0.7 and ages larger than 8~Gyr, or those of [Fe/H]=-0.4 and ages around 
6-8~Gyr, and also shows that the present fitting approach is more sensitive to the 
metallicity than to the age of the clusters. This application serves as an example 
of how to deal with the whole information contained in the spectra at once, 
irrespective of the spectral resolution involved.

As cited above, the crosscorrelation technique allowed us a rough but global evaluation 
of the goodness of our fits, however, for a further understanding of the nature of the 
present crosscorrelation results and for utilizing the advantages of the higher 
resolution, which selectively emphasizes certain individual features that give unique 
information (R94; Jones \& Worthey 1995), we have chosen to perform various key 
index-index diagrams of the type used in, e.g., Rose (1985), Rose \& Tripicco (1986) 
or R94. Thus, we first measured the Rose system of indices (see section 4.2) on the 
blue spectra of the two clusters, and on the synthetic SSP library prebroadened to 
match the resolution of the former ones. Fig.~6 shows various of these diagrams 
selected on the basis of their ability to separate the effects of the metallicity and 
the age. We see that the previous crosscorrelation results are robust but now much
strongly constrained, particularly, the first three plots indicate that the metallicity 
for the two clusters should be [Fe/H]$\sim$-0.7. The same diagrams also suggest ages 
in the range 10-15~Gyr. Our metallicity prediction for 47~Tuc is in agreement with Zinn 
(1985) and Webbink (1985), however our estimation for NGC~6624 is lower than the value 
corresponding to the first author but more similar to the one tabulated in the second 
paper. Also, from the line-strengths of the Lick indices given in Table~3 of Covino et 
al. (1995) or those of Table~3 of Cohen et al. (1998), we do not find significant 
differences in the measured values between these two clusters suggesting similar 
metallicities, in agreement with our result here.
Finally, in the 3888/3859 vs. H$\delta$/4045 plot of Fig.~6 we see that 
the observed values do not follow the model lines, showing the strong CN absorption 
feature for the two metal-rich clusters. This well known anomaly is present in both, 
their individual stellar spectra (e.g., Norris \& Freeman 1979) or in their integrated 
light (e.g., Rose \& Tripicco 1986). See also the discussion by R94 who built a synthetic 
spectrum for studying 47~Tuc. Tripicco \& Bell (1992) also built a CN-strong SSP high 
resolution spectrum, based on the oxygen-enhanced theoretical isochrones of VandenBerg 
(1992), and varied the Carbon and Nitrogen abundances to fit the 47~Tuc CN feature 
in a satisfactory way. Our results here are in good agreement with their predicted 
age and metallicity. 

Fig.~7 shows a 13~Gyr model spectrum of metallicity [Fe/H]=-0.7, conveniently 
smoothed, overplotted on the blue spectrum of 47~Tuc. This figure not only shows 
the quality of the fit, but also the quality of the flux-calibration of 
the model spectrum which is in excellent agreement with the observed cluster 
spectrum. As discussed above, the model spectrum, which is based on solar element 
ratios stellar spectra, is not able to fit the strong CN anomaly of this 
cluster but it is very sensitive in detecting such spectral peculiarities. 
Fig.~8 shows the normalized synthetic red spectrum (13~Gyr, [Fe/H]=-0.7) 
slightly broadened, overplotted on the observational spectrum of NGC~6624 at 
2.1\AA~FWHM. 

The present section not only served for comparing our new model library with metal-rich 
globular clusters, but also showed how easily this library can be applied to study them, 
by first adapting the synthetic spectra to the characteristics of the observational data 
and, then, performing the comparison either using the whole spectrum at once, or  
studying a set of well defined indicators at the highest spectral resolution allowed by 
the data. The flux-calibratioin of the data helps in making even easier and safer this 
study, as well as to extract most the information provided by wider features. We provided 
excellent fits which allow to infer the ages and metallicities of these two metal-rich
globular clusters. This study also showed the capacity of this 
new SSP spectral library to detect spectral anomalies in agreement with previous 
determinations.

\section{An application to an early-type galaxy}
Here we address the potential power of the new model predictions, as well as the
way in which they can be used, for analyzing the galaxy data. In particular it will 
be shown that we can adapt our theoretical spectra to the characteristics of the 
observational galaxy spectrum and then performing a comparison in a consistent
and very easy way. 
As an example, we have chosen an standard elliptical galaxy, NGC~3379, that we had 
studied in detail in Paper II by taking into account most of the spectroscopic
absorption features of the Lick system and the broad band colors from U to K. The full 
chemo-evolutionary version of our code (see Paper I) predicted a 
very small spread in metallicity for the central regions of this galaxy, and that 
the bulk of the stars were formed at the very early stages of the galactic 
evolution (at lower ages than $\sim$1~Gyr) (see Paper II for details). 
Therefore single-age, single-metallicity stellar populations are suitable to provide 
reasonable fits to this galaxy. In fact, in that paper we found an acceptable 
fit at 5$\arcsec$ of its center using an SSP of metallicity slightly larger 
than solar and $\sim$13~Gyr. These fits were found after excluding all the 
iron-dominated Lick features, because they were very weak when compared to most 
of the other Lick indices.

For studying this inner region of NGC~3379 (at $\sim$13$\%$ of its effective 
radius) using the new models, we first smoothen the synthetic spectra to 
match the velocity dispersion ($\sim235~Kms^{-1}$) obtained at this galactocentric
distance. 
In this way it will be no longer necessary to calculate any velocity dispersion 
correction factors to analyze correctly certain absorption features (e.g, 
Davies et al. 1993). Since our observed galaxy spectrum is not flux-calibrated, 
we performed once again a very careful continuum removal to ensure a correct
comparison, e.g., for using the crosscorrelation technique. The resulting peak 
heights are plotted in Fig.~9. One sees that the three models provide equally 
high values, but at slightly younger ages when increasing the metallicity from 
[Fe/H]=-0.4 to [Fe/H]=+0.2. 
This behavior differs from the ones that we obtained for the two globular 
clusters analyzed in section~5, where only the lowest metallicities produced the 
largest heights in an appreciable way. Fig.~10 shows the galaxy spectrum with a 
representative model spectrum for each metallicity overplotted. Each of the three 
model spectra provide reasonable fits for H$_{\beta}$, our most metal-rich model 
produces a reasonable Mg$_{b}$ feature at 5175\AA, but too strong iron features 
(e.g., Fe5015; Fe5270; Fe5335; Fe5406). The solar model spectrum almost fits the 
Mg$_{b}$ feature, but still provides strong iron-dominated features. On the other 
hand, the half a solar metallicity spectrum does not provide enough depth to the 
iron features and not at all to the Mg$_{b}$ feature. 
If we fix the age for the three metallicity models at, e.g., $\sim$15~Gyr 
(following Fig.~9), the half a solar model ([Fe/H]=-0.4) will produce slightly 
worse fits since the slightly younger age weakens the magnesium and the 
iron-dominated features. From this discussion it is now rather clear why different
models of different metallicities can provide equal approaches to the galaxy spectrum:
once H$_{\beta}$ is properly constrained, either we are able to fit the iron features 
(with a metallicity somewhat in the range -0.4$\leq$[Fe/H]$\leq$0) or the magnesium 
features (with a metallicity in the range 0$\leq$[Fe/H]$\leq$+0.2), but never
the two set of indices simultaneously, in agreement with our deep analysis performed 
in Paper II. Thus, the new model spectra were able to identify the galaxy 
spectral peculiarities such as differences in the element ratios. A great advantage 
here is that the analysis has been done in a very simple way, and in particular 
avoiding the tedious and difficult task to match the spectrum shape of the 
observational data to the instrumental response curve of any system of indices 
(e.g., Lick), and without calculating the required velocity dispersion correction 
factor for each individual feature. In the present approach, we have utilized all
the spectral information contained in the observational data, only limited by the
galaxy internal velocity dispersion. 

\section{Conclusions}

In this paper the evolutionary stellar population synthesis model of Vazdekis et al. 
(1996) is extended to predicts spectral energy distributions corresponding to SSP's 
at resolution 1.8\AA~FWHM for two narrow, but very important wavelength 
ranges, around 4000\AA~and 5000\AA. The new model uses as a database a subsample 
of 547 stars selected from the original Jones (1997) stellar spectral library, 
after a careful atmospheric parameters selection and removal of most of the 
peculiar stars (see also Vazdekis 1998). Therefore, this is the first time that 
an evolutionary model employs such an extensive empirical stellar spectral library, 
at such high resolution, for supporting its SED's predictions. 

A spectral library corresponding to simple old stellar populations with 
metallicities in the range $-0.7\leq[Fe/H]\leq+0.2$ is presented here, as well 
as an extensive discussion about the most popular system of absorption indices 
at intermediate resolution, the extended Lick system (Worthey et al. 1994 
and Worthey \& Ottaviani 1997), showing the difficulties for a proper 
transformation of the observational data to that system. Also, for the first 
time is shown the behavior of the system of indices of Rose (1994) at higher 
resolution as a function of the age and the metallicity. The model spectral 
library can be retrieved from the web homepage of the author 
(http://www.ioa.s.u-tokyo.ac.jp/\~{ }vazdekis/).

The newly synthesized model spectra can be used to analyze the observed galaxy 
spectrum in a safe and flexible way, allowing us to adapt the theoretical 
predictions to the nature of the data instead of proceeding in the opposite 
direction as, for example, we should do when transforming the observational data 
to perform an analysis using model predictions based on a particular 
instrumental dependent system of indices at a specific resolution(s) (e.g., the 
Lick system). The synthetic SSP library, with flux-calibrated spectral response, 
can be smoothed to the same resolution of the observations or to the measured 
internal galaxy velocity dispersion, allowing us to analyze the observed spectrum 
in its own system and therefore utilizing all the information contained in the data, 
at their higher spectral resolution. After performing this step, the entire 
observational spectrum can be compared at once, or the analysis can be done  
measuring a particular set of features in the two, the synthesized and the observed 
spectrum, rather than trying to correct the latter from broadening or instrumental 
effects, e.g., which characterize the system at which we desire to transform. 
For a more correct and easy analysis the spectra should be normalized or they
can be compared directly if the observational data are flux-calibrated. In fact, 
we strongly recommend to carry out this step (even before any normalization), 
which does not only places all the data at the same reference, but also for allowing 
to detect spectral peculiarities at larger scale, such as very strong Mg$_{2}$ or 
CN absorption features. 

The SSP model spectra were calibrated at relatively high resolution with two
standard metal-rich globular clusters, providing very good fits which not only
allowed to predict their metallicity and age, but also to being able 
to detect well known spectral peculiarities such as the CN anomaly. 
The library was also confronted to an standard early-type galaxy, NGC~3379, 
revealing its well known magnesium over iron overabundance (Vazdekis et al. 1997),
showing how appropriate are the new model predictions and the proposed methodology
of analysis to study the elemental ratios.
In fact, different models of different metallicities provide equal approaches 
to the galaxy spectrum: once H$_{\beta}$ is properly constrained, either we are 
able to fit the iron features (with a metallicity somewhat in the range 
-0.4$\leq$[Fe/H]$\leq$0) or the magnesium features (with a metallicity in the 
range 0$\leq$[Fe/H]$\leq$+0.2), but never the two set of indices simultaneously.

As mentioned above, the predicted spectra offer the possibility to perform
the study comparing any set of absorption indices (which falls in these 
spectral ranges) with the corresponding observed ones, after measuring them under 
the same conditions. An extensive set of tables and spectra corresponding to 
simple stellar populations with different model parameters, including the 
variation of the IMF, will be presented elsewhere. In that paper we explore 
the behavior of the different indicators as a function of the spectral
resolution, i.e., the galaxy velocity dispersion. Vazdekis \& Arimoto (1998) make 
use of the present SSP spectral library to redefine the H$\gamma$ 
index for increasing its unprecedent age discriminating power (Jones \& Worthey
1995), and for studying its behavior as a function of the resolution and other 
factors.

The stellar population studies would strongly benefit from expanding the present 
stellar database to those parametrical regions not well covered by stars.
Particularly, observations of metal-poor and metal-rich stars as well as 
very cool and hot stars are required to enlarge considerably the present range
of predictions for the stellar populations of galaxies. This not only will allow
to analyze a larger variety of galaxies, but also for performing more reliable
predictions from more complex models which include, e.g., chemical evolution.
Also, very important will be to extend the present wavelength coverage to 
account for the contributions from different stellar populations.

\acknowledgments
The author is indebted to L. Jones for making available his library, to 
V. Vansevi\^cius for providing reddening values for the stellar sample, to 
J. Rose and S. Covino for providing their globular cluster spectra and to 
N. Cardiel and J. Gorgas for providing with a set of their stellar spectra before 
publication. The author is specially grateful to N. Arimoto, V. Vansevi\^cius,
R. Peletier and the non anonymous referee, J. Rose, for very useful comments 
and lots of suggestions which greatly improved this paper. My special thanks 
goes to my wife Sonia for lots of help and patience. The author thanks 
the Japan Society for Promotion of Science for financial support.
This work was financially supported in part by a Grant-in-Aid for 
the Scientific Research (No.09640311) by the Japan Ministry of Education, 
Culture, Sports and Science.

\onecolumn
\hoffset -1.8truecm
\begin{table}
\tiny
\begin{center}
\begin{tabular}{r|rrrrrrrrrrrrrrrrr}
\hline\hline
Age&CN1&CN2&Ca4227&G&Fe4383&H$_{\beta}$&Fe5015&Mg$_{1}$&Mg$_{2}$&Mg$_{b}$&Fe5270&Fe5335&Fe5406
&H$\delta_{A}$&H$\gamma_{A}$&H$\delta_{F}$&H$\gamma_{F}$\\
\hline
\multicolumn{18}{c}{[Fe/H]=-0.7}\\
\hline
6.3&-0.044&-0.008&0.764&4.235&2.457&2.322&3.409&0.045&0.128&2.210&1.880&1.593&1.023&1.267&-2.087&1.718&0.460
 \\
7.0&-0.041&-0.005&0.788&4.390&2.571&2.220&3.431&0.048&0.132&2.271&1.904&1.625&1.046&0.968&-2.458&1.556&0.246
 \\
8.0&-0.036&-0.001&0.821&4.606&2.729&2.079&3.459&0.052&0.138&2.356&1.936&1.668&1.077&0.554&-2.973&1.331&-0.052
\\
9.0&-0.035&-0.001&0.841&4.686&2.775&2.037&3.440&0.053&0.141&2.399&1.942&1.673&1.080&0.435&-3.143&1.271&-0.152
\\
10.0&-0.034&-0.001&0.860&4.767&2.821&1.994&3.421&0.054&0.144&2.442&1.947&1.677&1.083&0.317&-3.313&1.210&-0.252\\
11.0&-0.036&-0.004&0.899&4.884&2.858&1.917&3.376&0.053&0.144&2.477&1.923&1.663&1.079&0.147&-3.576&1.099&-0.413\\
12.0&-0.029&0.003&0.923&5.022&2.965&1.821&3.410&0.054&0.146&2.498&1.968&1.710&1.108&-0.124&-3.912&0.955&-0.588\\
13.0&-0.026&0.006&0.913&5.086&2.988&1.827&3.419&0.055&0.147&2.517&1.969&1.706&1.102&-0.188&-3.997&0.930&-0.626\\
14.0&-0.025&0.007&0.942&5.209&3.095&1.775&3.463&0.057&0.152&2.594&2.008&1.739&1.127&-0.357&-4.251&0.846&-0.766\\
15.0&-0.028&0.004&0.953&5.289&3.077&1.754&3.431&0.057&0.151&2.574&1.983&1.715&1.116&-0.334&-4.319&0.840&-0.798\\
16.0&-0.026&0.005&0.970&5.393&3.172&1.731&3.482&0.059&0.154&2.599&2.019&1.747&1.137&-0.449&-4.506&0.785&-0.890\\
17.0&-0.019&0.013&0.990&5.577&3.337&1.708&3.645&0.063&0.161&2.694&2.088&1.803&1.167&-0.693&-4.816&0.677&-1.036\\
\multicolumn{18}{c}{[Fe/H]=-0.4}\\
\hline
3.2&-0.040&0.000&0.818&3.771&2.778&2.519&3.731&0.054&0.143&2.359&2.136&1.842&1.202&1.550&-1.492&1.962&0.890
  \\
4.0&-0.042&-0.003&0.844&3.934&2.821&2.490&3.720&0.054&0.145&2.390&2.134&1.830&1.193&1.378&-1.736&1.863&0.769
 \\
5.0&-0.036&0.001&0.902&4.218&3.010&2.369&3.843&0.058&0.152&2.496&2.191&1.887&1.229&0.927&-2.321&1.617&0.455
  \\
6.0&-0.028&0.008&0.980&4.558&3.297&2.209&3.944&0.063&0.163&2.660&2.281&1.972&1.285&0.335&-3.065&1.326&0.054
  \\
7.0&-0.022&0.013&1.034&4.805&3.496&2.083&4.027&0.066&0.169&2.761&2.343&2.034&1.327&-0.073&-3.617&1.121&-0.243
\\
8.0&-0.017&0.018&1.079&5.005&3.653&1.975&4.100&0.069&0.174&2.836&2.392&2.085&1.362&-0.392&-4.075&0.959&-0.490
\\
9.0&-0.015&0.020&1.130&5.147&3.782&1.903&4.141&0.072&0.180&2.949&2.434&2.127&1.392&-0.645&-4.414&0.844&-0.682
\\
10.0&-0.013&0.021&1.180&5.290&3.911&1.832&4.182&0.074&0.186&3.063&2.476&2.168&1.423&-0.897&-4.753&0.728&-0.875\\
11.0&-0.012&0.022&1.198&5.364&3.980&1.786&4.176&0.077&0.189&3.070&2.500&2.185&1.432&-1.021&-4.934&0.657&-0.979\\
12.0&-0.009&0.025&1.254&5.480&4.128&1.729&4.203&0.079&0.195&3.188&2.547&2.232&1.466&-1.289&-5.243&0.543&-1.157\\
13.0&-0.006&0.028&1.283&5.540&4.209&1.693&4.250&0.080&0.198&3.228&2.578&2.253&1.488&-1.457&-5.438&0.453&-1.270\\
14.0&-0.003&0.031&1.314&5.586&4.288&1.656&4.253&0.082&0.201&3.282&2.607&2.283&1.514&-1.627&-5.603&0.383&-1.363\\
15.0&-0.003&0.030&1.337&5.622&4.335&1.632&4.259&0.083&0.204&3.329&2.625&2.294&1.517&-1.715&-5.711&0.338&-1.428\\
16.0&-0.004&0.029&1.348&5.656&4.360&1.609&4.256&0.084&0.205&3.350&2.632&2.315&1.535&-1.753&-5.796&0.317&-1.478\\
17.0&-0.001&0.032&1.351&5.688&4.401&1.589&4.273&0.086&0.207&3.371&2.647&2.337&1.550&-1.817&-5.874&0.282&-1.520\\
\multicolumn{18}{c}{[Fe/H]=0.0}\\
\hline
1.6&-0.043&0.000&0.804&3.262&2.839&3.115&4.368&0.049&0.140&2.205&2.330&2.031&1.306&2.379&-0.364&2.473&1.711
  \\
2.0&-0.033&0.008&0.894&3.728&3.159&2.879&4.504&0.054&0.150&2.413&2.409&2.100&1.366&1.678&-1.344&2.101&1.187
  \\
2.5&-0.019&0.020&0.993&4.249&3.589&2.575&4.607&0.062&0.163&2.624&2.516&2.205&1.447&0.778&-2.573&1.631&0.525
  \\
3.0&-0.011&0.026&1.082&4.607&3.930&2.433&4.738&0.067&0.174&2.814&2.631&2.304&1.512&0.163&-3.362&1.348&0.109
  \\
4.0&0.008&0.045&1.236&5.110&4.580&2.182&5.002&0.075&0.196&3.156&2.846&2.490&1.651&-0.898&-4.680&0.881&-0.598
 \\
5.0&0.007&0.043&1.298&5.306&4.732&2.088&4.938&0.078&0.202&3.268&2.875&2.517&1.671&-1.222&-5.136&0.714&-0.865
 \\
6.0&0.009&0.044&1.360&5.442&4.883&2.008&4.928&0.082&0.210&3.390&2.921&2.563&1.703&-1.537&-5.510&0.573&-1.077
 \\
7.0&0.014&0.048&1.413&5.572&5.056&1.918&4.958&0.086&0.217&3.500&2.978&2.613&1.742&-1.889&-5.906&0.416&-1.304
 \\
8.0&0.020&0.054&1.465&5.696&5.240&1.826&5.007&0.089&0.223&3.604&3.040&2.666&1.784&-2.256&-6.306&0.253&-1.534
 \\
9.0&0.026&0.061&1.545&5.748&5.432&1.768&5.083&0.093&0.232&3.719&3.117&2.740&1.838&-2.589&-6.608&0.121&-1.693
 \\
10.0&0.032&0.067&1.625&5.799&5.623&1.710&5.159&0.097&0.240&3.834&3.195&2.813&1.892&-2.922&-6.910&-0.011&-1.853\\
11.0&0.039&0.074&1.706&5.835&5.809&1.642&5.241&0.102&0.248&3.955&3.265&2.882&1.946&-3.248&-7.187&-0.138&-2.003\\
12.0&0.045&0.081&1.765&5.890&5.968&1.596&5.305&0.106&0.255&3.999&3.329&2.945&1.995&-3.490&-7.443&-0.239&-2.134\\
13.0&0.044&0.079&1.777&5.894&5.969&1.573&5.251&0.105&0.255&4.057&3.327&2.942&1.990&-3.592&-7.497&-0.291&-2.184\\
14.0&0.047&0.084&1.823&5.905&6.044&1.530&5.273&0.108&0.259&4.098&3.360&2.974&2.013&-3.719&-7.619&-0.345&-2.249\\
15.0&0.050&0.087&1.854&5.918&6.116&1.501&5.303&0.111&0.262&4.104&3.387&3.003&2.036&-3.806&-7.717&-0.382&-2.294\\
16.0&0.051&0.087&1.868&5.930&6.122&1.453&5.300&0.112&0.262&4.071&3.390&3.009&2.046&-3.824&-7.783&-0.413&-2.332\\
17.0&0.054&0.091&1.936&5.923&6.224&1.408&5.338&0.116&0.269&4.156&3.432&3.053&2.080&-3.994&-7.921&-0.480&-2.400\\
\multicolumn{18}{c}{[Fe/H]=+0.2}\\
\hline
1.6&-0.023&0.018&0.968&4.024&3.622&2.933&4.892&0.058&0.162&2.532&2.629&2.311&1.485&1.444&-1.894&2.030&0.989
  \\
2.0&-0.011&0.029&1.060&4.393&4.016&2.751&5.141&0.066&0.175&2.755&2.743&2.433&1.569&0.789&-2.790&1.711&0.523
  \\
2.5&0.000&0.038&1.169&4.820&4.441&2.547&5.267&0.071&0.188&3.000&2.866&2.539&1.642&0.001&-3.811&1.344&-0.024
  \\
3.0&0.009&0.046&1.231&5.033&4.714&2.408&5.345&0.077&0.199&3.166&2.950&2.615&1.697&-0.500&-4.401&1.114&-0.355
 \\
4.0&0.040&0.077&1.402&5.430&5.385&2.115&5.613&0.093&0.226&3.528&3.161&2.825&1.864&-1.653&-5.739&0.615&-1.092
 \\
5.0&0.049&0.086&1.451&5.634&5.619&2.019&5.658&0.098&0.235&3.676&3.229&2.873&1.906&-2.249&-6.289&0.349&-1.420
 \\
6.0&0.051&0.088&1.518&5.715&5.748&1.960&5.650&0.102&0.242&3.767&3.273&2.907&1.944&-2.536&-6.561&0.226&-1.574
 \\
7.0&0.059&0.096&1.580&5.787&5.918&1.891&5.683&0.106&0.250&3.882&3.334&2.963&1.990&-2.894&-6.878&0.085&-1.753
 \\
8.0&0.069&0.107&1.640&5.854&6.106&1.818&5.735&0.111&0.258&4.006&3.401&3.028&2.040&-3.282&-7.213&-0.063&-1.941
\\
9.0&0.078&0.116&1.718&5.892&6.277&1.744&5.780&0.116&0.267&4.106&3.462&3.092&2.090&-3.594&-7.499&-0.188&-2.091
\\
10.0&0.086&0.125&1.797&5.930&6.447&1.670&5.825&0.121&0.275&4.206&3.523&3.156&2.140&-3.906&-7.785&-0.313&-2.241\\
11.0&0.088&0.127&1.835&5.955&6.501&1.625&5.851&0.124&0.280&4.250&3.547&3.178&2.161&-4.041&-7.913&-0.376&-2.318\\
12.0&0.092&0.131&1.908&5.964&6.631&1.574&5.880&0.129&0.288&4.356&3.601&3.227&2.199&-4.299&-8.103&-0.469&-2.425\\
13.0&0.095&0.134&1.977&5.972&6.738&1.528&5.900&0.132&0.293&4.430&3.638&3.268&2.226&-4.477&-8.262&-0.533&-2.505\\
14.0&0.102&0.142&2.067&5.992&6.916&1.473&5.962&0.136&0.302&4.533&3.697&3.324&2.272&-4.759&-8.498&-0.631&-2.616\\
15.0&0.104&0.145&2.079&5.984&6.938&1.455&5.955&0.138&0.303&4.524&3.709&3.332&2.281&-4.798&-8.530&-0.646&-2.638\\
16.0&0.110&0.151&2.115&5.978&7.025&1.422&5.984&0.140&0.307&4.583&3.743&3.360&2.306&-4.956&-8.637&-0.704&-2.695\\
17.0&0.117&0.158&2.119&5.985&7.062&1.399&5.990&0.142&0.310&4.621&3.763&3.372&2.318&-5.079&-8.714&-0.747&-2.752\\
\end{tabular}
\end{center}
\caption{The Lick indices measured on the obtained synthetic spectra for SSP's
of different metallicities and ages (in Gyr) and Salpeter IMF, after smoothing 
as indicated in Table~2. For the models of [Fe/H]=+0.2 we refer the reader to 
the last paragraph of section~3.  All the indices are expressed in EW, except the 
CN1, CN2, Mg$_{1}$ and Mg$_{2}$ which are given in magnitudes. Colors and 
other predictions can be found in Paper I.}
\end{table}

\hoffset -1.8truecm
\begin{table}
\tiny
\begin{center}
\begin{tabular}{r|rrrrrrrrrrrrrrrrr}
\hline\hline
[Fe/H]&CN1&CN2&Ca4227&G&Fe4383&H$_{\beta}$&Fe5015&Mg$_{1}$&Mg$_{2}$&Mg$_{b}$&Fe5270&Fe5335&Fe5406
&H$\delta_{A}$&H$\gamma_{A}$&H$\delta_{F}$&H$\gamma_{F}$\\
\hline
-0.7&-0.003&-0.001&-0.09&0.53&-0.21&0.01&-0.58& 0.002&-0.005&-0.08&-0.13&
0.01&0.05&0.10&-0.11&0.15&-0.12\\
-0.4& 0.003& 0.006&-0.09&0.33&-0.13&0.13&-0.42&-0.002&-0.008&-0.04&-0.03& 0.04&0.07&0.26&
0.03&0.19& 0.02\\
 0.0&-0.001& 0.003& 0.00&0.17&-0.10&0.20&-0.35&-0.015&-0.022&-0.16& 0.06&
 0.00&0.06&0.16&-0.29&0.15&-0.06\\
+0.2& 0.004& 0.010&-0.08&0.09&-0.19&0.30&-0.24&-0.019&-0.027&-0.22&
0.09&-0.09&0.02&0.24&-0.28&0.24&-0.03\\
\end{tabular}
\end{center}
\caption{Transformation factors between the model indices obtained on the basis 
of the empirical fitting functions (Lick system instrumental response curve, i.e., Paper I) 
and those measured on the new model SSP spectra (nearly flux calibrated) after convolving 
with a gaussian of $\sigma$=5.7 pixels (0.6\AA/pix) for the blue and $\sigma$=5.9 pixels for
the red  
($\sim$255 and $\sim$215~Kms$^{-1}$ respectively). The tabulated numbers are obtained
averaging the differences between the two predictions for all the ages presented in Table~1.
A negative value denotes that the index measured on the synthetic spectra is lower}
\end{table}

\hoffset -1cm

\begin{figure}
\plotone{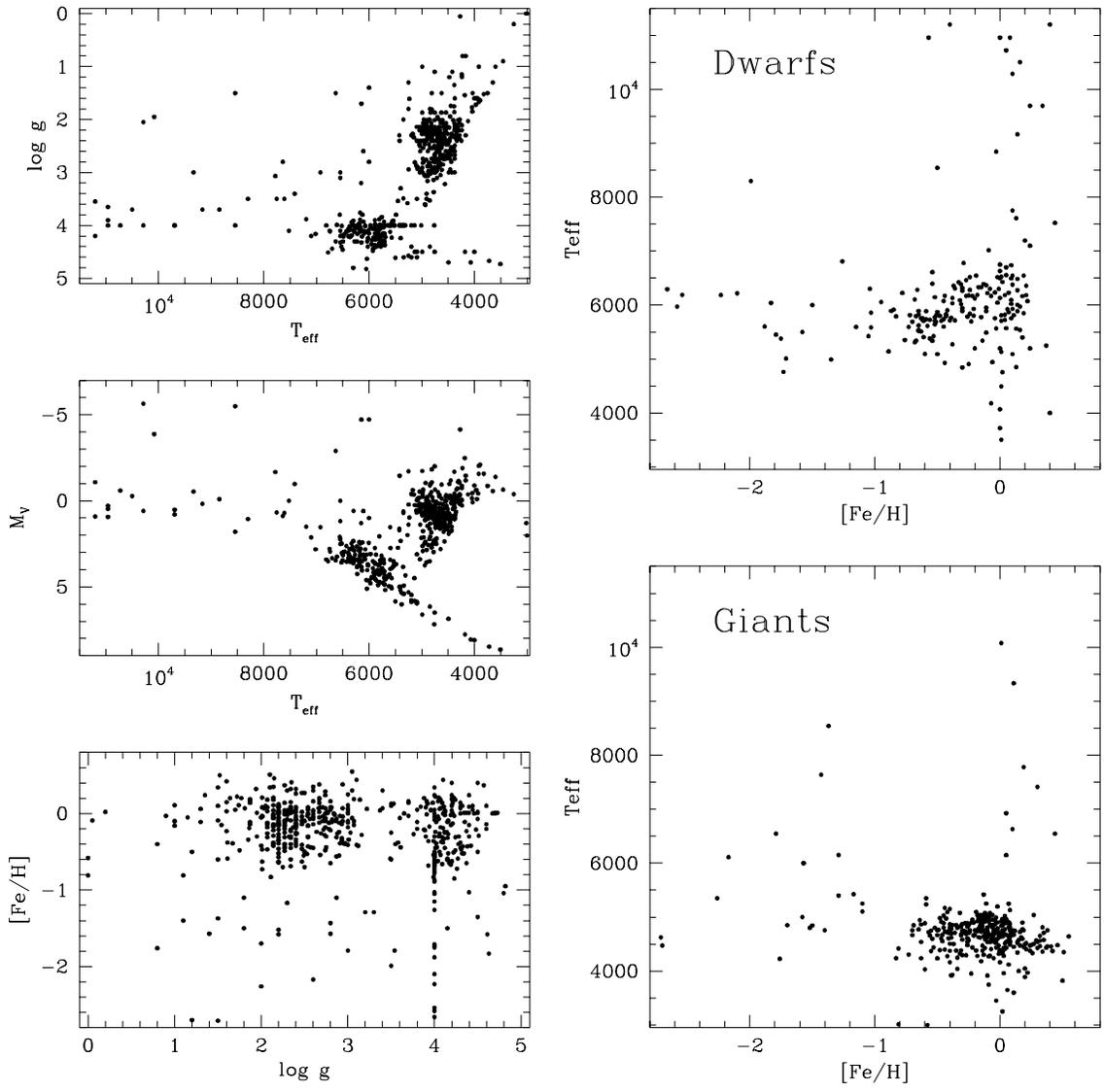}
\caption{The adopted atmospheric stellar parameters for the final subsample 
composed of 547 stars}
\end{figure}
\begin{figure}
\plotone{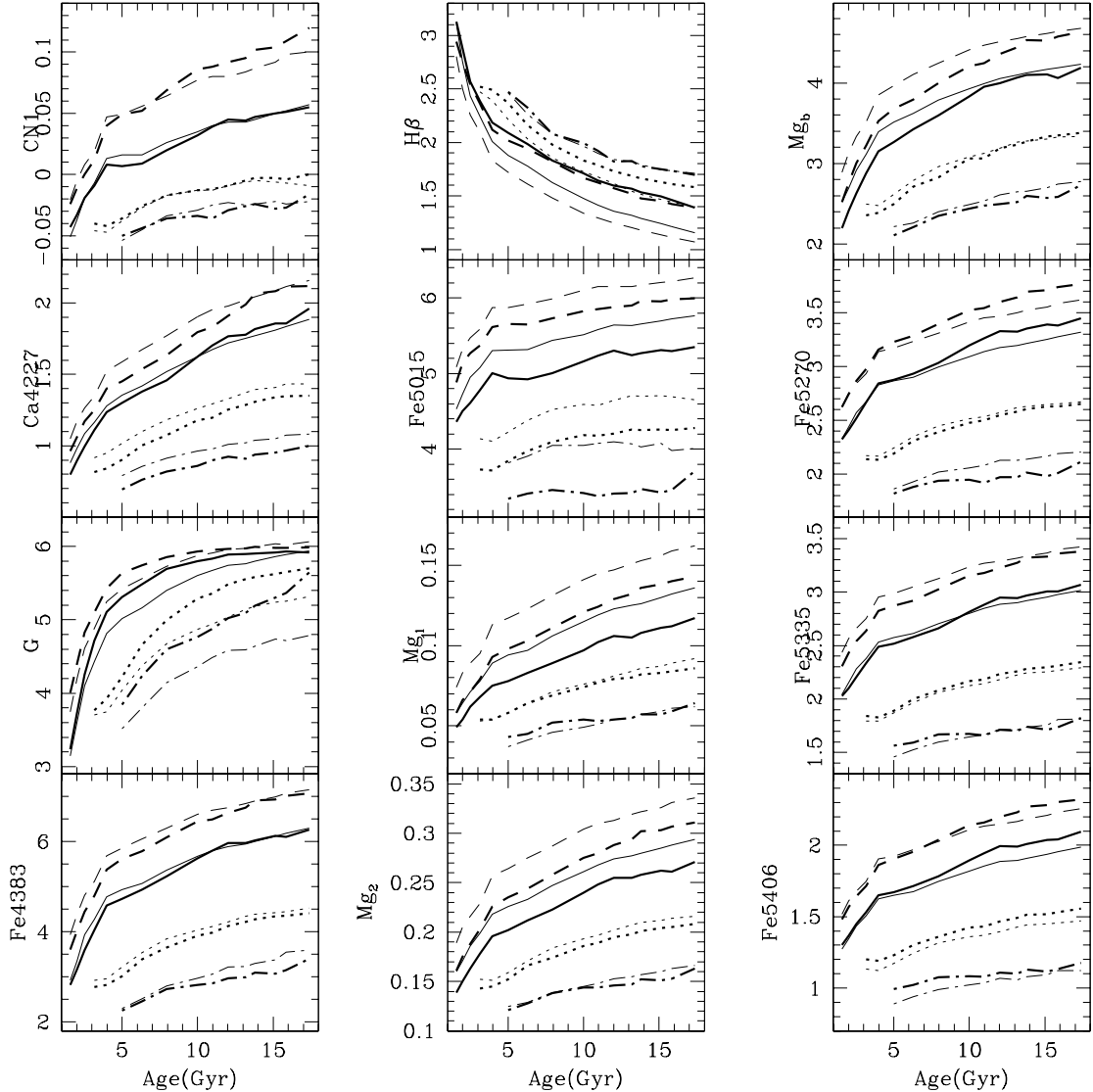}
\caption{The Lick system of indices. The thick lines represent the measurements 
obtained on the synthetic model spectra after smoothing to the Lick resolution: 
the high resolution synthetic spectra were convolved with a gaussian so that the 
strong resolution sensitive indices, the Ca4227 in the blue and the Fe5335 in the
red, were fully matched to the predictions obtained on the basis of the empirical 
fitting functions for solar metallicity. Dot dashed lines mean models of 
[Fe/H]=-0.7, dotted lines are models of [Fe/H]=-0.4, solid lines are models of 
solar metallicity and the dashed lines correspond to the predictions for 
[Fe/H]=+0.2 (using the approach discussed in the latest paragraph of section~3). 
Finally, the thin lines are the Paper I model predictions based on the empirical 
fitting functions of Worthey et al. (1994). The thin dashed lines
represents the predictions for [Fe/H]=+0.2, after interpolating between the 
values given in Paper I for solar and 2.5 times solar metallicity predictions}
\end{figure}
\begin{figure}
\plotone{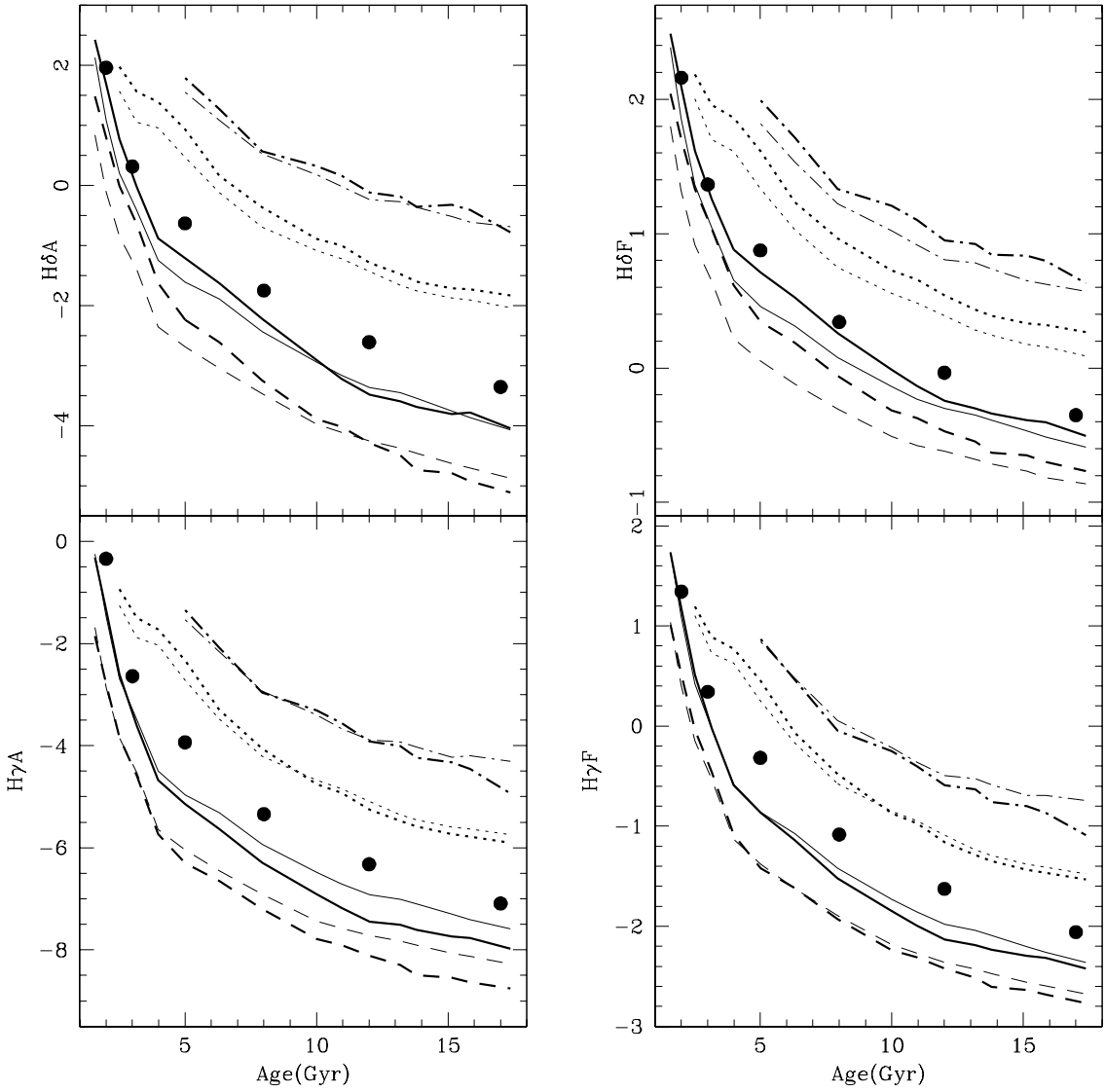}
\caption{The system of indices of WO97. The lines have the same meaning as in 
Fig.~2. Big solid dots correspond to the solar metallicity model predictions 
of these authors}
\end{figure}
\begin{figure}
\plotone{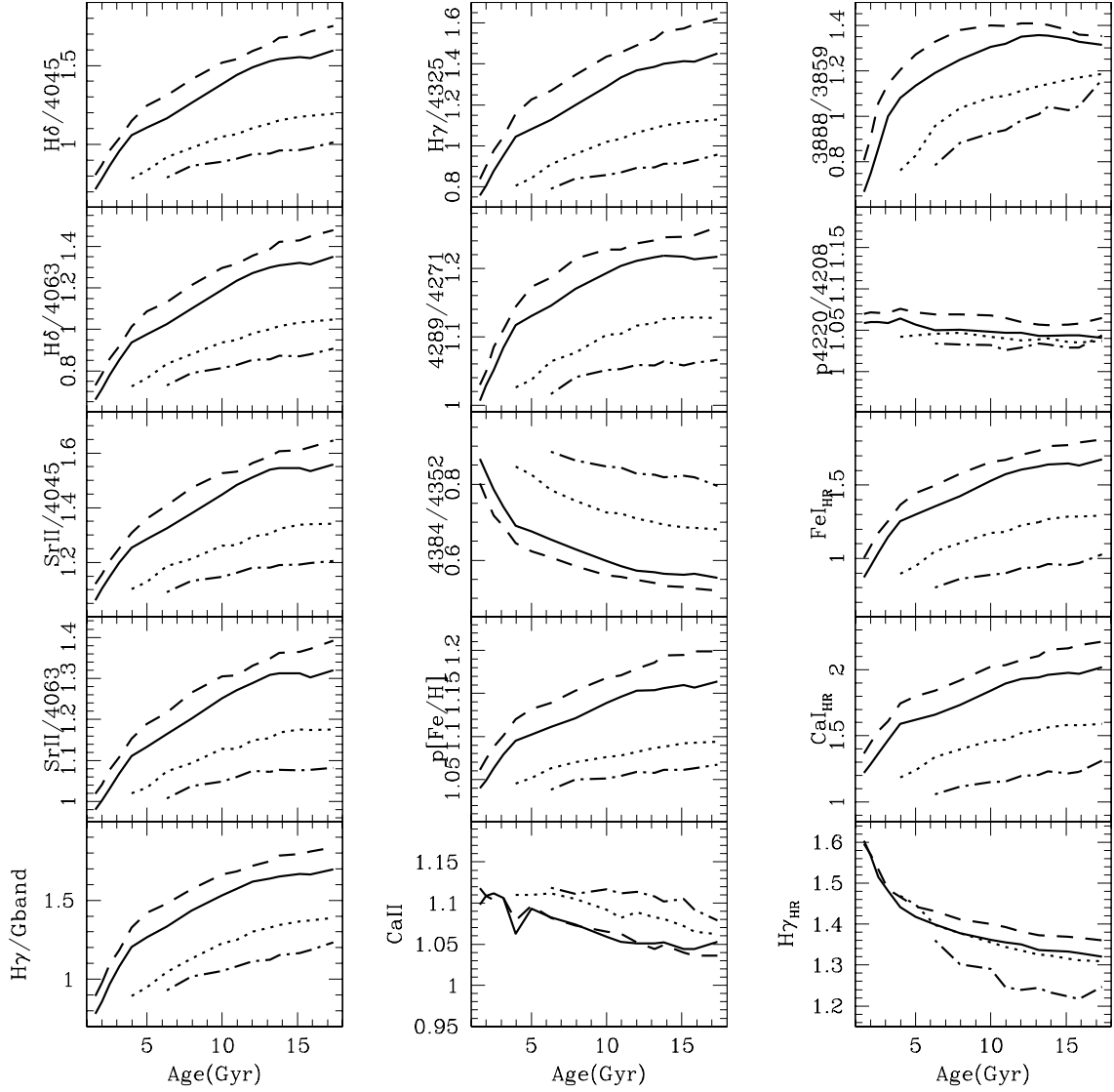}
\caption{The system of indices of R94. The values were measured on the same 
synthetic spectra as in Fig.~2 and Fig.~3 but without performing any degradation
of the resolution (1.8\AA~ FWHM). The line types have the same meaning as in 
these two figures. Here we warn that the resolution plays an important role and we 
will address this problem elsewhere}
\end{figure}
\begin{figure}
\plotone{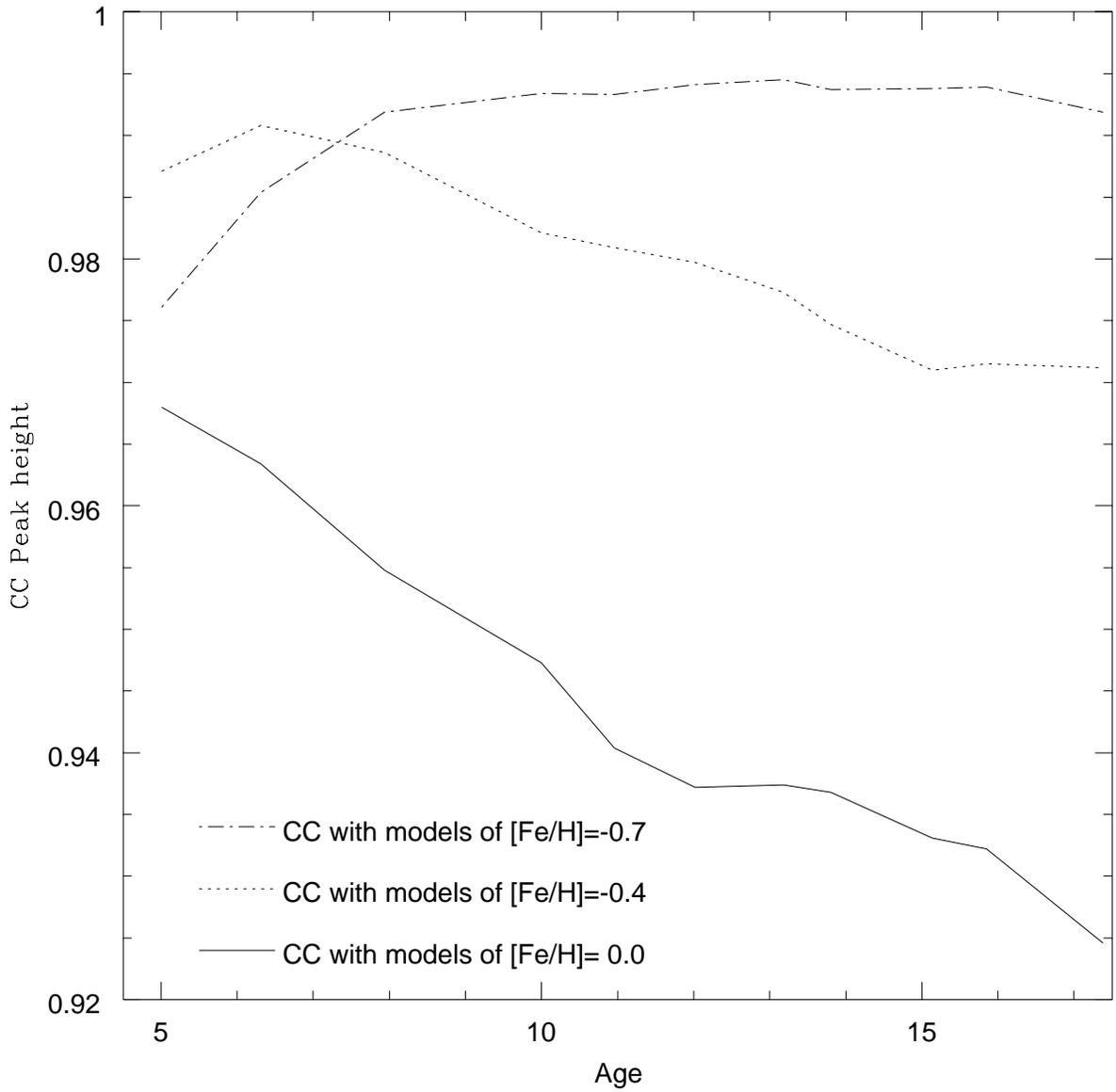}
\caption{Peak height resulting from the crosscorrelation of the 47~Tuc
integrated 
blue spectrum with different SSP model spectra of various metallicities and ages 
(see the text for details). Notice that old populations of 
[Fe/H]=-0.7 provide the best fits}
\end{figure}
\begin{figure}
\plotone{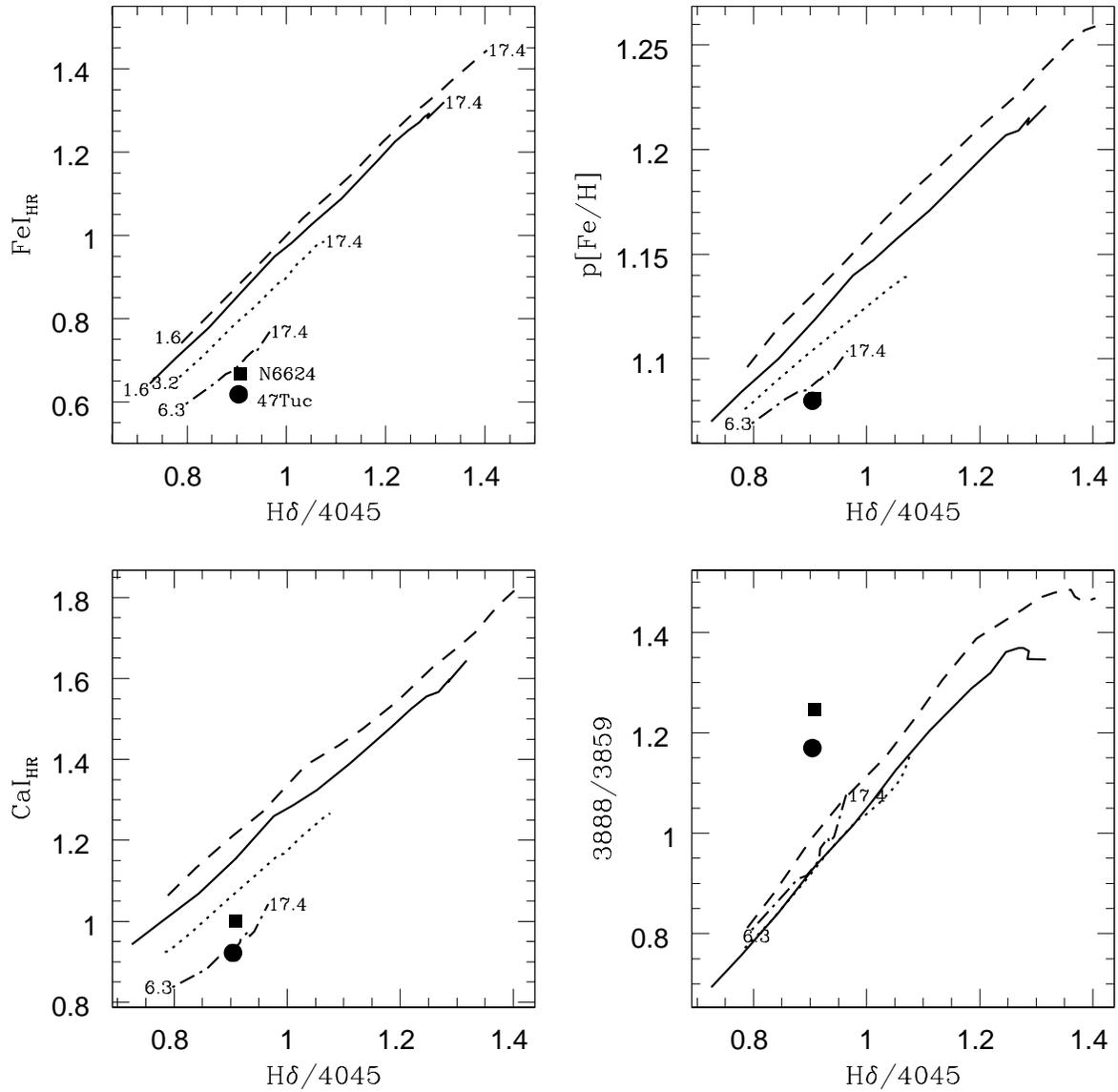}
\caption{Plot of various index-index diagrams of the Rose system. The model lines
have the same meaning as in Fig.~2. We have marked for each metallicity the lowest and 
highest ages (in Gyr)}
\end{figure}
\begin{figure}
\plotone{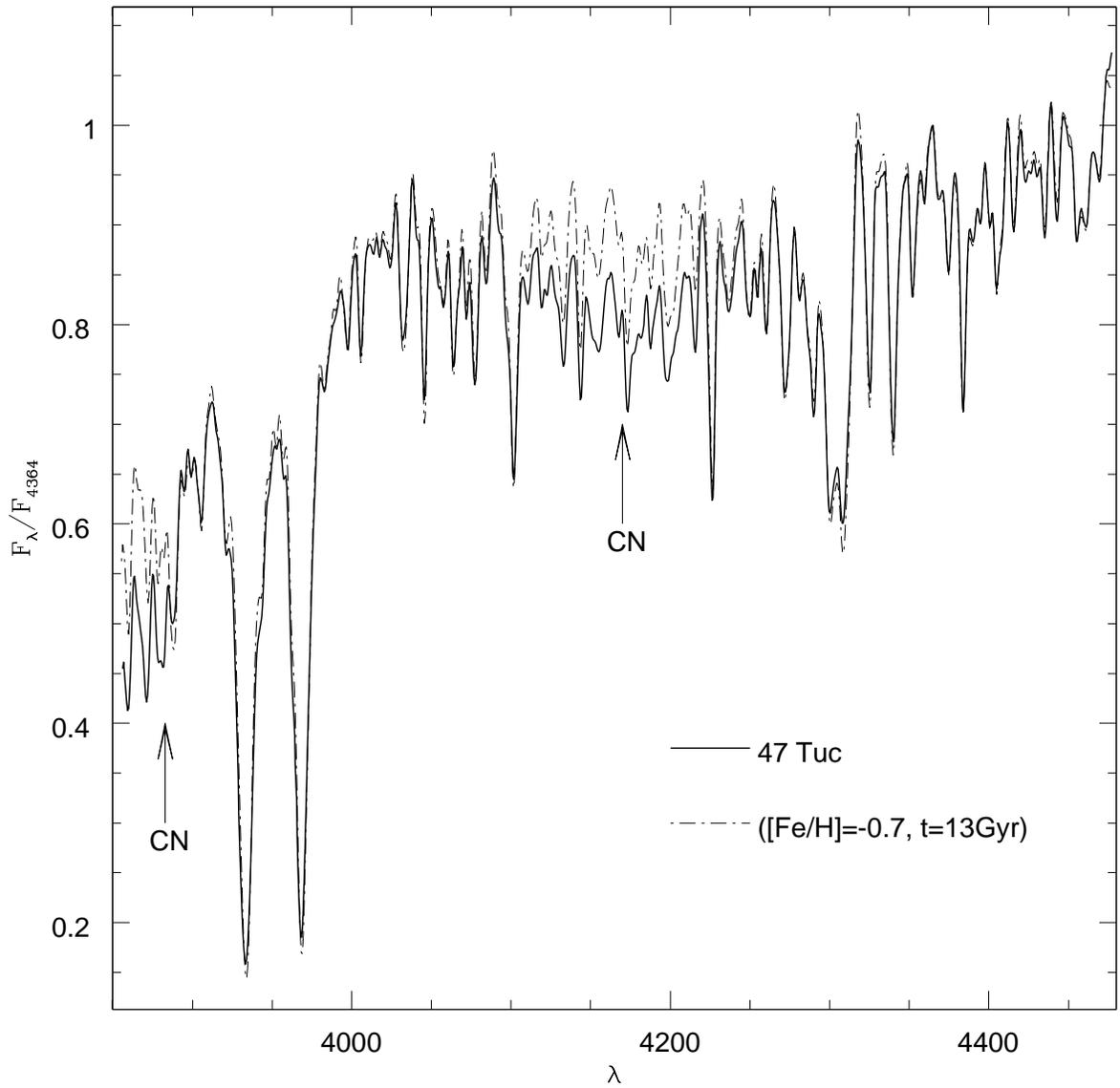}
\caption{The 47 Tuc blue spectrum and a representative good fit model spectrum 
selected on the basis of Fig.~6. We have marked the well known CN strong bands 
of this globular cluster which has been detected by the model}
\end{figure}
\begin{figure}
\plotone{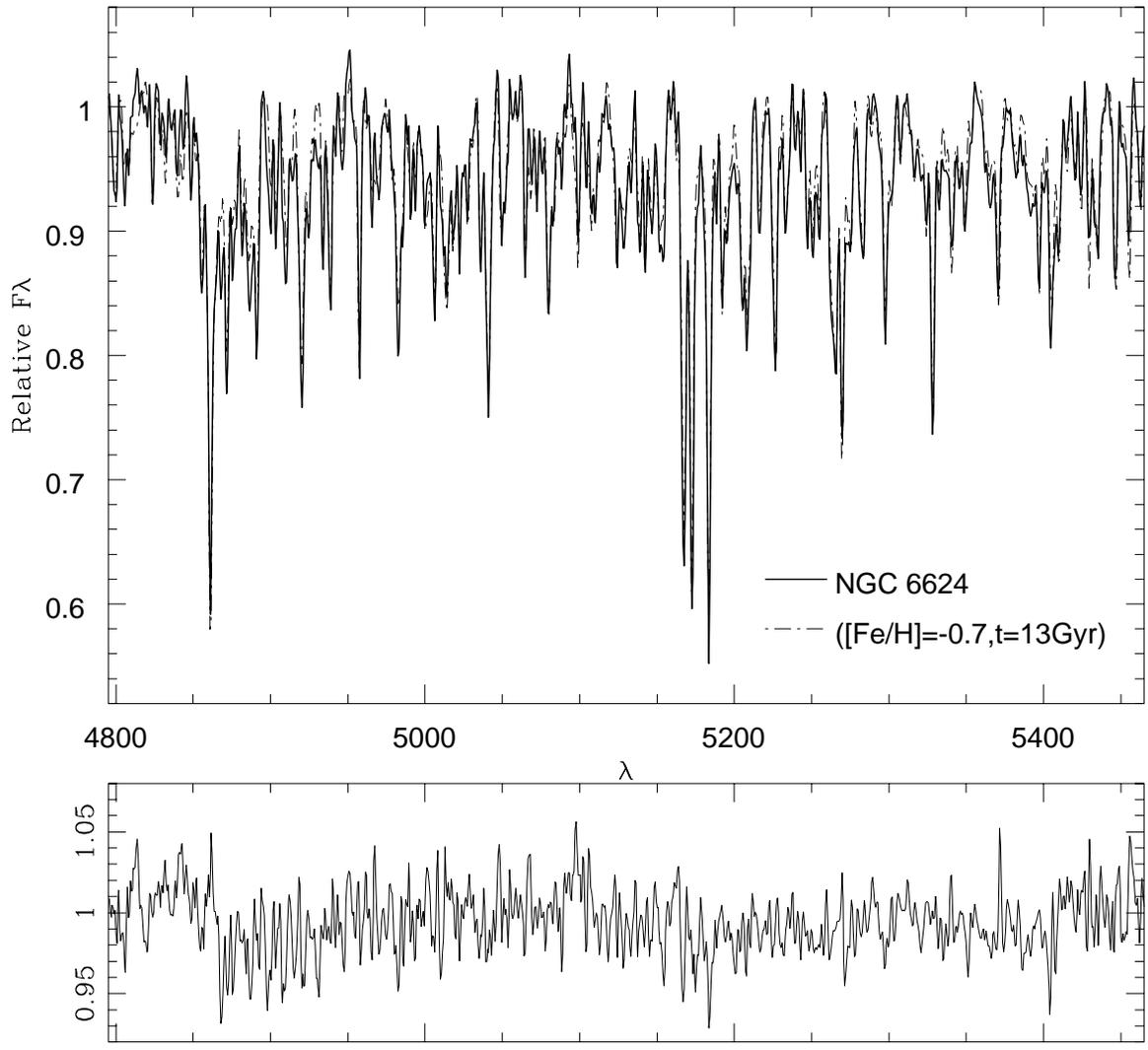}
\caption{The normalized NGC~6624 red spectrum and a representative good fit model 
spectrum at $\sim$2.1\AA~FWHM spectral resolution. Also plotted is the ratio
between the cluster and the model spectra}
\end{figure}
\begin{figure}
\plotone{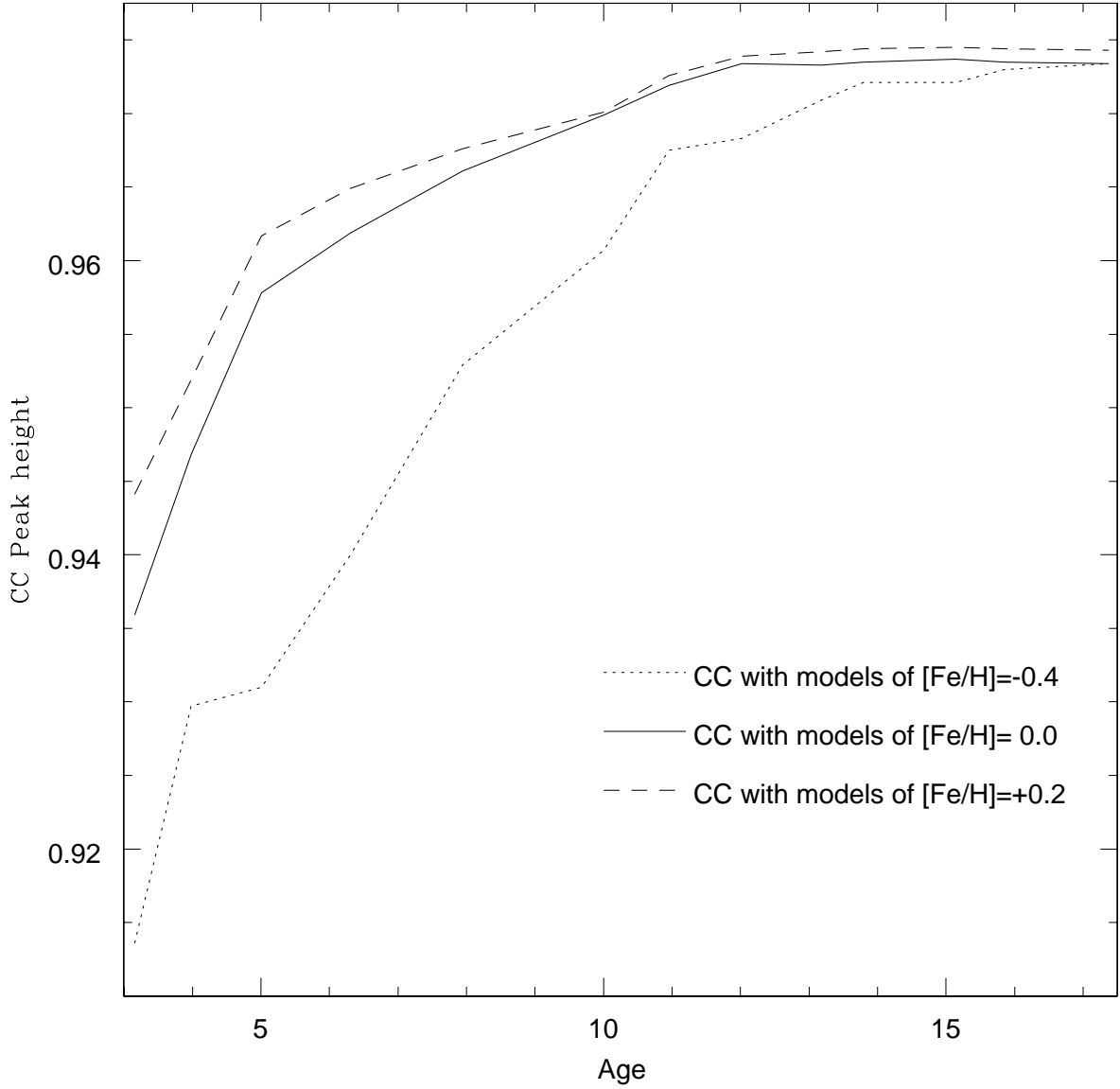}
\caption{Peak heights resulting from the crosscorrelation of the NGC~3379 galaxy 
red spectrum at $\sim$13$\%$ of its effective radius with different model spectra 
of various metallicities and ages. Notice that models of different metallicities 
provide very similar peak heights varying slightly the age}
\end{figure}
\begin{figure}
\plotone{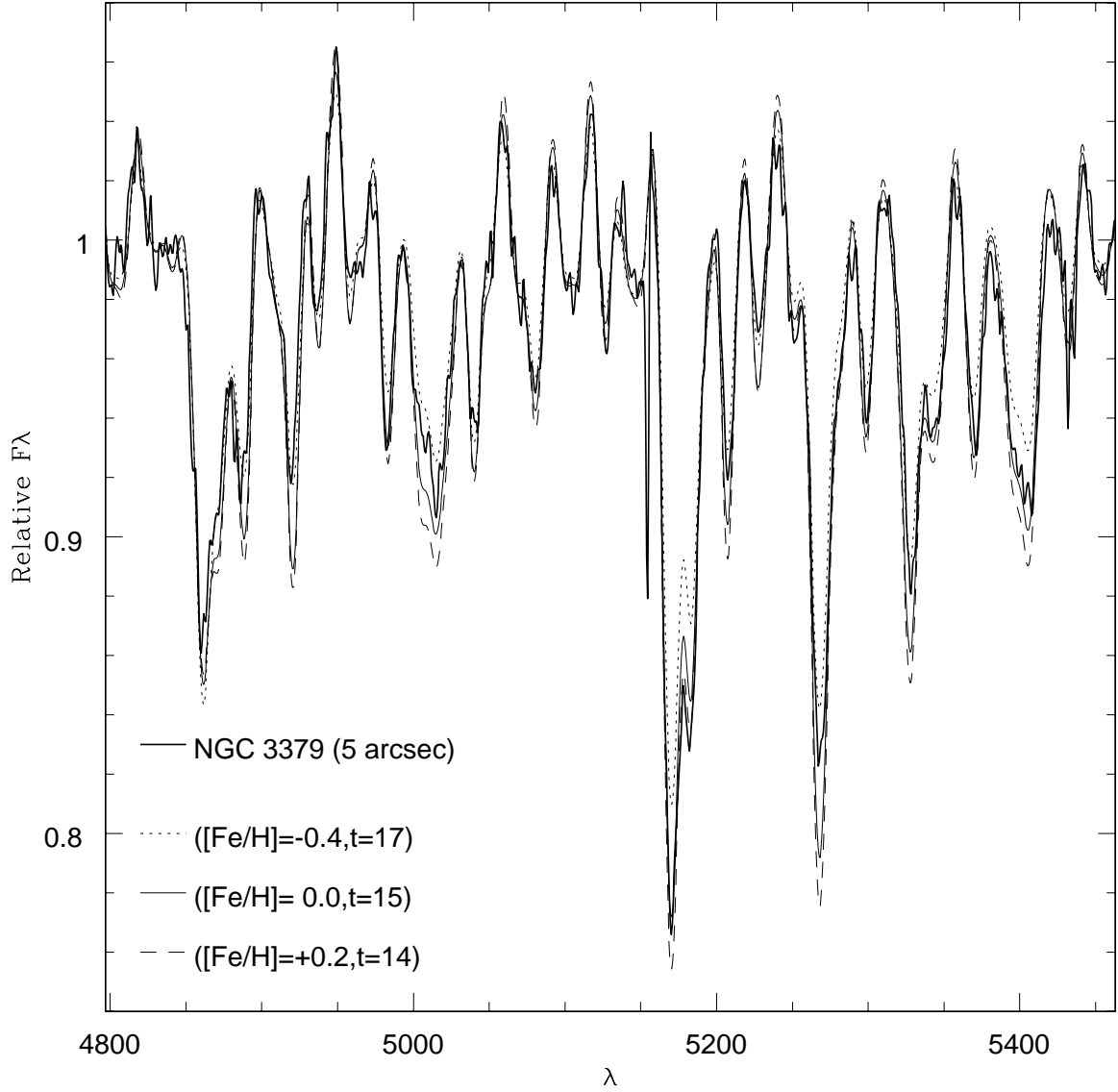}
\caption{The normalized NGC~3379 red spectrum. Overplotted are three model spectra
selected on the basis of Fig.~9. Notice that all of them give good H$_{\beta}$
($\sim$4860\AA) fits but no one provides acceptable fits to the magnesium 
($\sim$5175\AA) and iron features (e.g., Fe5015; Fe5270; Fe5335; Fe5406) 
simultaneously}
\end{figure}

\end{document}